\documentclass[a4paper,fleqn,usenatbib]{mnras}

\usepackage[T1]{fontenc}
\usepackage{ae,aecompl}

\usepackage{graphicx}	
\usepackage{amsmath}	
\usepackage{amssymb}	
\usepackage{textcomp}

\usepackage{mathptmx}
\usepackage{txfonts}

\title[Observations of CIZA J2242.8+5301 with the SRT]{Observations of the galaxy cluster CIZA J2242.8+5301 with the Sardinia Radio Telescope}

\author[F. Loi et al.]{
F. Loi,$^{1,2}$\thanks{E-mail: floi@oa-cagliari.inaf.it}
M. Murgia,$^{1}$
F. Govoni,$^{1}$
V. Vacca,$^{1} $
L. Feretti,$^{3}$
G. Giovannini,$^{3,4}$
E. Carretti,$^{1} $
\newauthor F. Gastaldello,$^{5}$
M. Girardi,$^{6,7}$
F. Vazza,$^{3,8}$
R. Concu,$^{1}$
A. Melis,$^{1}$
R. Paladino,$^{3}$
S. Poppi,$^{1}$
\newauthor G. Valente,$^{1,9}$
W. Boschin,$^{10,11,12}$
T.E. Clarke,$^{13}$ 
S. Colafrancesco,$^{14}$
T. En{\ss}lin,$^{15}$
\newauthor C. Ferrari,$^{16}$
F. de Gasperin,$^{17}$
L. Gregorini,$^{3} $
M. Johnston-Hollitt,$^{18,19}$
H. Junklewitz,$^{20}$
\newauthor E. Orr{\`u},$^{21}$
P. Parma,$^{3}$
R. Perley,$^{22}$
G.B Taylor,$^{23}$  
\\
$^{1}$INAF-Osservatorio Astronomico di Cagliari, Via della Scienza 5, I-09047 Selargius (CA), Italy \\
$^{2}$Dipartimento di Fisica, University of Cagliari, Strada Prov.le Monserrato-Sestu Km 0.700,I-09042 Monserrato (CA), Italy          \\      
$^{3}$INAF - Istituto di Radioastronomia, Via Gobetti 101, I--40129 Bologna, Italy \\
$^{4}$Dip. di Fisica e Astronomia, Universit\`a degli Studi Bologna, Viale Berti Pichat 6/2, I--40127 Bologna, Italy \\
$^{5}$INAF - IASF Milano, Via Bassini 15, I-20133 Milano, Italy  \\
$^{6}$Dip. di Fisica dell'Universit\`a degli Studi di Trieste - Sezione di Astronomia, via Tiepolo 11, I-34143 Trieste, Italy  \\
$^{7}$INAF - Osservatorio Astronomico di Trieste, via Tiepolo 11, I-34143 Trieste, Italy \\
$^{8}$Hamburger Sternwarte, Universit\"at Hamburg, Gojenbergsweg 112, 21029, Hamburg, Germany \\
$^{9}$Agenzia Spaziale Italiana (ASI), Roma \\
$^{10}$Fundaci\'on G. Galilei - INAF TNG, Rambla J. A. Fern\'andez P\'erez 7, E-38712 Bre\~na Baja (La Palma), Spain \\
$^{11}$Instituto de Astrof\'{\i}sica de Canarias, C/V\'{\i}a L\'actea s/n, E-38205 La Laguna (Tenerife), Spain\\
$^{12}$Dep. de Astrof\'{\i}sica, Univ. de La Laguna, Av. del Astrof\'{\i}sico Francisco S\'anchez s/n, E-38205 La Laguna (Tenerife), Spain\\
$^{13}$Naval Research Laboratory, Washington, District of Columbia 20375, USA  \\
$^{14}$School of Physics, University of the Witwatersrand, Private Bag 3, 2050, Johannesburg, South Africa \\
$^{15}$Max Planck Institut f\"{u}r Astrophysik, Karl-Schwarzschild-Str.1, 85740 Garching, Germany \\
$^{16}$Laboratoire Lagrange, UCA, OCA, CNRS, Blvd de l'Observatoire, CS 34229, 06304 Nice cedex 4, France \\
$^{17}$University of Leiden, Rapenburg 70, 2311 EZ Leiden, the Netherlands \\
$^{18}$Peripety Scientific Ltd., PO Box11355 Manners Street, Wellington, 6142, New Zealand\\  
$^{19}$School of Chemical $\&$ Physical Sciences, Victoria University of Wellington, PO Box 600, Wellington, 6140, New Zealand \\ 
$^{20}$Argelander-Institut f\"{u}r Astronomie, Auf dem H\"{u}gel 71 D-53121 Bonn, Germany \\
$^{21}$ASTRON, the Netherlands Institute for Radio Astronomy, Postbus 2, 7990 AA, Dwingeloo, The Netherlands \\
$^{22}$National Radio Astronomy Observatory, P.O. Box O, Socorro, NM 87801, USA \\
$^{23}$Department of Physics and Astronomy, University of New Mexico, Albuquerque NM, 87131, USA \\
}

\date{Accepted XXX. Received YYY; in original form ZZZ}

\pubyear{2017}

\begin{document}
\label{firstpage}
\pagerange{\pageref{firstpage}--\pageref{lastpage}}
\maketitle

\clearpage
\begin{abstract}
We observed the galaxy cluster CIZA J2242.8+5301 with the Sardinia Radio Telescope to provide 
new constraints on its spectral properties at high frequency.
We conducted observations in three frequency bands centred at 1.4\,GHz, 6.6\,GHz and 19\,GHz, 
resulting in beam resolutions of 14$^{\prime}$, 2.9$^{\prime}$ and 1$^{\prime}$  respectively. 
These single-dish data were also combined with archival interferometric observations at 1.4 and 1.7\,GHz. 
From the combined images, we measured a flux density of ${\rm S_{1.4GHz}=(158.3\pm9.6)\,mJy}$ for the central radio halo and
${\rm S_{1.4GHz}=(126\pm8)\,mJy}$ and ${\rm S_{1.4GHz}=(11.7\pm0.7)\,mJy}$ for the northern and the southern relic respectively.
After the spectral modelling of the discrete sources, we measured at 6.6\,GHz ${\rm S_{6.6GHz}=(17.1\pm1.2)\,mJy}$ and ${\rm S_{6.6GHz}=(0.6\pm0.3)\,mJy}$ for the northern and southern relic respectively. 
Assuming simple diffusive shock acceleration, we interpret measurements of the northern relic with a continuous injection model
represented by a broken power-law. 
This yields an injection spectral index ${\rm \alpha_{inj}=0.7\pm0.1}$ and a Mach number ${\rm M=3.3\pm0.9}$, consistent with recent X-ray estimates.
Unlike other studies of the same object, no significant steepening of the relic radio emission is seen in data up to 8.35\,GHz.
By fitting the southern relic spectrum with a simple power-law (${\rm S_{\nu}\propto\nu^{-\alpha}}$) we obtained a spectral index ${\rm \alpha\approx1.9}$
corresponding to a Mach number (${\rm M\approx1.8}$) in agreement with X-ray estimates.
Finally, we evaluated the rotation measure of the northern relic at 6.6\,GHz.
These results provide new insights on the magnetic structure of the relic, but further observations are needed to clarify the nature of the observed Faraday rotation.
\end{abstract}




\section{Introduction}
In the standard scenario of hierarchical formation of the Universe large-scale structures form and grow through merger events 
involving dark matter dominated clumps like galaxy clusters and galaxy groups.
An example of these processes are gravitational driven collisions between two sub-clusters of galaxies that end up with the formation of a more massive galaxy cluster.
Such merging events represent the most spectacular and energetic processes since the Big Bang: they can release enormous amounts of energy in the intracluster medium (ICM), 
as much as ${\rm \gtrsim10^{64}\,erg}$ \citep{sarazin}.

In these environments some galaxy clusters show diffuse synchrotron radio sources called radio haloes and relics \citep[e.g.,][]{feretti12} 
found at the centre and at the periphery of the cluster respectively.
These sources are not associated with specific optical counterparts so they reveal the existence of a non-thermal component in the ICM
made up of relativistic particles and magnetic fields spread over the cluster volume \citep[e.g.,][]{cartay}.
Radio haloes and relics have a typical size of ${\rm \sim1\,Mpc}$, low surface brightness (${\rm \sim 0.1-1\,\mu Jy/arcsec^{2}}$ at 1.4\,GHz) and are steep-spectrum (${\rm \alpha \gtrsim1}$).
Relics are elongated arc-like sources usually oriented perpendicularly to the merger axis.
While radio haloes are generally observed to be unpolarized there are a few cases where it has been possible to detect filaments of polarized emission 
(e.g. in A2255, \citet{a2255}; MACS J0717.5+3745, \citet{macs}; A523, \citet{a523}). 
In contrast, relics exhibit a high degrees of polarization (${\rm \sim20-30\%}$) indicating that either 
the magnetic field in these sources is extremely ordered or it has been confined in a thin layer.

From the study of relics one can make a step forward in understanding the physics of the ICM.
Observations are consistent with the idea that relics trace shock waves occurring in merger events \citep{roe},
where shocks are identified as sharp discontinuities in the temperature and surface brightness distributions inferred from X-ray observations of galaxy clusters.
In particular, \citet{en98} proposed a diffusive shock acceleration mechanism \citep[DSA,][]{dsa} to explain the re-acceleration of relativistic particles in relics.
This mechanism does not put constraints on the nature of the injected relativistic particles.
In this respect, recent observations are consistent with a scenario in which old electrons from nearby 
radio galaxies are re-accelerated through the DSA mechanism \citep[e.g.,][]{bonafede14,van17}. 
However, a single acceleration of thermal electrons by shocks or a combination of acceleration and re-acceleration have been shown to produce quite similar results
in a few specific cases \citep[e.g.,][]{kang}.
Radio observations of relics are also important to understand the evolution of large scale magnetic fields.
Shock waves amplify the fields to ${\rm \mu}$G-levels and align the magnetic field with the shock plane, as predicted by the aforementioned models and
observed in terms of strong linearly-polarized synchrotron emission associated with relics.
Relics are usually observed at low frequencies with interferometers.
However, due to the fact that interferometers are only sensitive to angular scales
according to their minimum baseline length, single-dish telescopes are essential for the study of diffuse radio sources, especially at high frequencies \citep{emerson}.
There are many examples of single-dish telescopes being used to investigate diffuse radio relics and haloes in literature \citep{farn,carretti,maja}.

The Sardinia Radio Telescope\footnote{www.srt.inaf.it} (SRT) is a new 64-m single-dish radio telescope, 
designed to operate in the frequency range 0.3-116\,GHz, and currently working between 0.3 and 26\,GHz \citep{bolli,scicom}.
It has recently been successful in spectro-polarimetric observations of galaxy clusters \citep{matte16}. 
In this paper we present the SRT observations of the galaxy cluster CIZA J2242.8+5301 (hereafter CJ2242). 
We observed CJ2242 with the SRT at 1.4\,GHz, 6.6\,GHz and 19\,GHz as part of of the early science project SRT Multi-frequency observations of galaxy clusters
(SMOG, project code S0001, PI M. Murgia).
The SMOG program consists of wide-band and wide-field spectral-polarimetric observations of a sample of galaxy clusters \citep{a194}. 
The aim of the project is to improve our knowledge of the non-thermal components (relativistic particles and magnetic fields) of the ICM on large scale
and to shed light on the interplay between these components and the life-cycles of cluster radio galaxies.
This can be done through the comparison of SRT observations with radio observations at higher resolution and at different frequencies, and with observations in the 
mm, sub-mm, optical and X-ray.

The galaxy cluster CJ2242 has been included in the SMOG sample because it is an interesting case of system hosting a faint central halo and a double relic system \citep{van10}
and it is located at a relatively low redshift of z=0.1921 \citep{koc}. 
The diffuse radio emission easily exceeds 15$^{\prime}$, 
making single-dish observations with the SRT essential to measure the emission at all relevant angular scales.

The paper is organized as follows.
In Section 2, we briefly illustrate the current knowledge about this cluster.
In Section 3, we describe the details of the observations, the data reduction and the imaging of CJ2242 with the L-band, C-band and K-band receivers of the SRT. 
Total intensity images are shown in Section 4, where we include the combination of single-dish and interferometric maps at 1.4\,GHz for the galaxy cluster.
We also estimate the flux density of the diffuse and discrete sources and their spectral behaviour.
In Section 5 we present the linearly-polarized emission and the application of the rotation measure synthesis technique to C-Band observations.
Finally, the conclusions are presented in Section 6.\\
Throughout this paper we assume a ${\rm\Lambda}$CDM cosmology with ${\rm H_0=71\,km \cdot s^{-1} Mpc^{-1}}$, ${\rm\Omega_m=0.27}$, ${\rm\Omega_{\Lambda}=0.73}$. 
At the redshift of CJ2242, 1$^{\prime}$ corresponds to 189.9\,kpc.

\section{The galaxy cluster CIZA J2242.8+5301}
\label{sect:ciza}
The galaxy cluster CJ2242 was discovered with X-rays by \citet{koc}
during a survey aiming at finding new cluster candidates at low Galactic latitudes (l=104.19 degrees, b=$-$5.11 degrees).
They measured its redshift (z=0.1921) and a luminosity (${\rm L_X=4.0 \times 10^{44}\,erg s^{-1}}$) in the energy band 0.1$-$2.4\,keV.
Its disturbed elongated X-ray morphology suggested that this could be an example of cluster that underwent a merger event.
\citet{van10} confirmed this scenario using observations taken with the Westerbork Synthesis Radio Telescope (WSRT), the Giant Metrewave Radio Telescope (GMRT) and the Very Large Array (VLA).
In their work they presented the discovery of a faint central halo and a double relic system with a northern relic having a length of $\sim$2\,Mpc, a width of $\sim$55\,kpc and 
located 1.5 Mpc from the centre of the cluster. 
The radio morphology, together with the spectral index gradient of the northern relic toward the cluster centre, 
have been interpreted as signatures of acceleration and spectral aging of relativistic electrons due to the passage of a shock.
Assuming the DSA model they found a Mach number ${\rm M\sim 4.6}$ for the northern relic. \\
A reasonable scenario for this galaxy cluster 
is a collision in the plane of the sky (within 10$^{\circ}$) between two sub-clusters with a mass ratio of 2:1 (confirmed by \citealt{dmshear} using gravitational lensing) 
and an impact parameter $\lesssim$400\,kpc.
The core passage of the sub-clusters likely happened about 1 Gyr ago \citep{van11}. \\
Association of the relics with shock fronts has been confirmed 
from Suzaku observations - \citet{aka} measured a drop in temperature at the position of the northern relic, while the jump in the X-rays surface brightness was not detected, probably because of resolution limits.
However, even with the highest spatial resolution X-ray data available \citep[i.e. the Chandra data,][]{ogrean14}, the surface brightness discontinuity is still not detected.
This aspect is very peculiar since at these high Mach numbers (M$\sim$2-4) the discontinuity should be remarkable, as is the case for El Gordo \citep{botteon} or A665 \citep{dasadia}. 
Projection effects can be invoked but the optical analysis by \citep{dawson15} points to the systems being in a nearly plane-of-the-sky merger configuration.
Evidence for shock compression east of the southern relic has been observed with XMM-Newton by \citet{ogrean}.
Recent estimates \citep{aka2015} of temperature drops from Suzaku observations suggest a Mach number M$\sim$2.7 and M$\sim$1.7 for the northern and the southern relics respectively.\\
The tension between Mach numbers for the northern relic inferred from radio and X-ray data was the first hint for physics beyond the DSA model.
As a result there were several attempts to find a consistent model for all observations of CJ2242 using numerical simulations \citep{donnert17}.  
However, new measurements of the radio flux injection index resulted in a Mach number M$\sim$2.9 \citep{stroe2014}, in better agreement with the X-ray value inferred from the temperature jump detected by Suzaku.
In recent years CJ2242 has been observed over several radio frequency bands.
In particular, \citet{stroe2016} presented a study of the northern relic from 150 MHz to 30\,GHz. 
They highlighted the steepening of the integrated spectrum beyond 2.5\,GHz (from ${\rm \alpha}$=0.90 to ${\rm \alpha}$=1.77), claiming that the simple standard relic scenario could not hold for this galaxy cluster. 
Several models have been proposed to explain the steepening
such as a time variable magnetic field in the relic area \citep{donnert}, or an additional population of fossil electrons that has broken out from the shock \citep{kang}.
However, interpretations based on the currently available high frequency data (at 16 and 30\,GHz) have to be taken with care because of their large uncertainties.
These measurements are made from interferometric observations that could have lost a significant fraction of the total flux and 
can also be noticeably affected by the Sunyaev$-$Zel\textasciigrave dovich effect \citep[see][]{basu}.
These aspects have been taken into account in a recent work where the relic has been studied with single-dish observations conducted with the Effelsberg telescope \citep{maja}. 
They inferred a relic spectral index ${\rm \alpha =(0.90\pm 0.04)}$ between 150 MHz and 8.35\,GHz and suggest that models describing the origin of relics
have to include effects beyond the DSA mechanism that requires ${\rm \alpha>1}$, in order to physically model the relic.
Similar findings have been reported for the cluster A2256 by \citet{trasatti}.\\
From this picture, it is clear how new, accurate high frequency observations and estimates of the flux density of the northern relic are important to constrain the physical 
scenario of CJ2242.

\begin{center}    
\begin{table*}\centering
\caption{Details of the observations of the galaxy cluster CIZA J2242.8+5301 at the SRT in L and C-bands, acquired during the SMOG program. 
  Columns report from left to right: 
  the date, the receiver, the frequency range in\,GHz, the beam full-width-half-maximum (FWHM), the field of view (FOV), the number of on-the-fly (OTF) maps carried out, 
  the calibrators used for the data reduction, the time spent on source (TOS).}      
\label{tab:obs}
\begin{tabular}{c c c c c c c c c c}     
     \hline
     Date & Receiver & Freq.$_{\rm [GHz]}$ & FWHM$_{[^\prime]}$ & FOV & OTF scan axis & Calibrators & TOS$_{\rm[h]}$ \\
     \hline
     08 Jul 2016 & L-Band & 1.3-1.8 & 14 & 3$^{\circ}\times3^{\circ}$ & 2$\times$RA+1$\times$DEC & 3C147 & 2.2\\
     12 Jul 2016 & L-Band & 1.3-1.8 & 14 & 3$^{\circ}\times3^{\circ}$ & 1$\times$RA+2$\times$DEC & 3C147 & 2.2\\
     \hline
     06 Feb 2016 & C-Band & 6.0-7.2 & 2.9 & 1$^{\circ}\times1^{\circ}$ & 2$\times$RA+2$\times$DEC & 3C286, 3C84 & 2.3\\
     01 Jun 2016 & C-Band & 6.0-7.2 & 2.9 &  1$^{\circ}\times1^{\circ}$ & 2$\times$RA+1$\times$DEC & 3C48, 3C84 & 2.5\\
     24 Jun 2016 & C-Band & 6.0-7.2 & 2.9 &  30$^{\prime}\times30^{\prime}$& 9$\times$RA+9$\times$DEC & 3C138, 3C48, 3C84 & 4.2\\
     26 Jun 2016 & C-Band & 6.0-7.2 & 2.9 & 30$^{\prime}\times30^{\prime}$ & 9$\times$RA+9$\times$DEC & 3C138, 3C48, 3C84 & 5\\
    \hline
\end{tabular}
\end{table*}   
\end{center}
\begin{center}    
\begin{table*}\centering
\caption{Details of the observations of some discrete sources, in the field of view of the galaxy cluster CIZA J2242.8+5301, at 19\,GHz with a FWHM of 1 $^{\prime}$. 
  Columns report from left to right: the source name, the source J2000 coordinates, the field of view (FOV) of the scans, the number of the cross scans, the calibrators used for the data reduction.}
\label{tab:cs}
\begin{tabular}{c c c c c c}    
    \hline
    Source & RA [$^h$ $^m$ $^s$]& DEC [$^{\circ}$ $^{\prime}$ $^{\prime\prime}$] & FOV & Cross Scans & Calibrators \\
    \hline
    A & 22 43 38.027 & +53 09 19.47 & 6$^{\prime}$x6$^{\prime}$ & 20 & 3C147 \\
    B & 22 42 44.666 & +53 08 04.98 & 6$^{\prime}$x6$^{\prime}$ & 21 & 3C147 \\
    C & 22 43 17.995 & +53 07 19.84 & 6$^{\prime}$x6$^{\prime}$ & 22 & 3C147 \\
    D & 22 42 48.004 & +53 05 34.99 & 6$^{\prime}$x6$^{\prime}$ & 23 & 3C147 \\
    E & 22 42 53.000 & +53 04 50.00 & 6$^{\prime}$x6$^{\prime}$ & 11 & 3C147 \\
    G & 22 42 51.338 & +53 00 35.00 & 6$^{\prime}$x6$^{\prime}$ & 23 & 3C147 \\
    H & 22 42 04.820 & +52 59 34.39 & 6$^{\prime}$x6$^{\prime}$ & 11 & 3C147 \\
   \hline
\end{tabular}
\end{table*}
\end{center}
\section{Observations and data reduction}
We observed the galaxy cluster CJ2242 at the SRT in three frequency ranges: L-band (1.3-1.8\,GHz), C-band (6-7.2\,GHz) and K-band (18-20\,GHz).
L-band and C-band observations were centred at the J2000 coordinates RA 22$^h$ 42$^m$ 53.0$^s$ and DEC +53$^{\circ}$ 01$^{\prime}$ 05$^{\prime\prime}$,
while K-band observations were centred on individual galaxies of the system.
The details of the observations are reported in Table \ref{tab:obs} (L, C-band observations) and Table \ref{tab:cs} (K-band observations).\\
We acquired spectral-polarimetric data in full-Stokes parameters with the 
SArdinia Roach2-based Digital Architecture for Radio Astronomy back-end (SARDARA, Melis et al. in prep.) for the L and C-band observations, 
while for the K-band observations only total intensity continuum observations were performed.
For the data reduction and the imaging we used the proprietary software package Single-dish Spectral-polarimetry Software \citep[SCUBE,][]{matte16}.
In the following we describe, for each band, the observational set up and the procedure adopted for the data reduction and imaging.
\subsection*{L-Band}
We observed an area of 3$^{\circ}\times3^{\circ}$
with the entire 1.3$-$1.8\,GHz band of the L-band receiver. We used the SARDARA back-end configuration with 1500 MHz bandwidth 
(the only option available to cover the entire 500 MHz band of the receiver) and 16384 channels of 92 kHz each.
The beam FWHM is about 14$^{\prime}$ at a frequency of 1.55\,GHz.
We performed several On-The-Fly (OTF) maps in equatorial coordinates along the two orthogonal directions of RA and DEC.
The telescope scanning speed was set to 6$^{\prime}$/s and the scans were separated by 3.5$^{\prime}$ to properly sample the SRT beam. 
We recorded the data stream by sampling at 10 spectra per second thereby producing individual samples separated by 36$^{\prime\prime}$ along the scanning direction.\\
For our purposes, we calibrated only the total intensity. Band-pass and flux density calibration were performed by observing 3C\,147 assuming the flux density scale of \citet{perley}. 
Those frequency windows affected by persistent radio-frequency interference (RFI) were flagged by hand. 
We also applied an automatic flagging procedure to excise the large amount of RFI randomly spread in frequency and time.
The flagged data were then used to repeat the baseline subtraction, bandpass, and flux density calibration. 
At these frequencies, the gain-elevation curve can be assumed to be flat and therefore we did not apply any correction.\\
We subtracted the baseline of the OTF maps of CJ2242, scan by scan, by fitting 10$\%$ of the data at the beginning and at the end of each scan.
We projected the data in a regular 3-dimensional grid with a spatial resolution of 180$^{\prime\prime}$/pixel.  We then applied the automatic flag procedure on the target
and we repeated the baseline subtraction and the projection.
In total, we discarded about 30\% of the data. 
The total intensity image of CJ2242 is obtained by stacking all the calibrated OTF maps.
In the combination the individual image cubes were averaged and de-striped by mixing their stationary wavelet transform (SWT) coefficients. 
Coefficients on a spatial scale below 2 pixels were omitted in order to improve the signal-to-noise ratio without degrading the resolution. 
For further details see \citet{matte16}.
\subsection*{C-Band}
We decided to make a shallow and a deep map, over a field of view of 1$^{\circ}\times1^{\circ}$ and 30$^{\prime}\times30^{\prime}$, respectively.
We used a 1.2\,GHz bandwidth centred at 6.6\,GHz.
We acquired full-Stokes parameters in a bandwidth of 1500 MHz with the SARDARA back-end, with 1024 channels of 1.46 MHz each.
The beam FWHM at 6.6\,GHz is 2.9$^{\prime}$.
We performed several OTF maps setting a spacing between the scans of 0.7$^{\prime}$ using
a scan rate of 6$^{\prime}$/s for the shallow map and 3$^{\prime}$/s for the deep map.
We acquired 33 spectra per second, therefore, on the sky, the spatial separation between individual samples along the scanning direction was 10.9$^{\prime\prime}$ and 5.45$^{\prime\prime}$, 
for the shallow and the deep maps respectively.\\
We calibrated the band-pass and the flux density by observing 3C\,286, 3C\,48 or 3C\,138, depending on availability for each observation,
assuming the flux density scale of \citet{perley}.
We removed RFI observed in a cold part of the sky and we repeated the calibration and RFI-flagging procedure until all the obvious RFI was removed.
Then, we applied the gain-elevation curve correction to take into account the gain variations with elevation due to the telescope structure's gravitational stress change.
For the polarization we first corrected the delay between right and left polarization of the receiver by using 3C 286, 3C 138 or 3C48.
Next we used 3C 84 to correct the instrumental polarization, and finally we corrected the absolute position of the polarization angle using 3C 286, 3C 138 or 3C48.\\
We subtracted the baseline of each scan of the OTF map of CJ2242.
To take into account the presence of bright sources at the edge of the images we decided to refine the baseline removal from each image cube using a mask. 
The mask has been obtained from the observation of CJ2242 at 1.4\,GHz with the NRAO VLA Sky Survey \citep[NVSS,][]{condon}: 
after a convolution of the map with the SRT beam we blanked all the pixels where the signal was larger than 1${\rm\sigma}$ (${\rm\sigma}$=0.5 mJy/beam), in order to keep just the noise regions.
We proceeded by fitting the baseline of our images with a second order polynomial, excluding the blanked regions in the mask. 
In this way, we could evaluate and subtract the noise more efficiently.
We flagged the remaining RFI and then we repeated the baseline removal. 
The fraction of flagged data is $\sim30\%$.
We projected the data in a regular 3-dimensional grid with a spatial resolution of 42$^{\prime\prime}$/pixel.
Afterwards, we stacked together all the RA-DEC scans to obtain full-Stokes I, U, and Q cubes.
In the combination the individual image cubes were averaged and de-stripped by mixing their SWT coefficients. 
Coefficients on a spatial scale below 2 pixels were omitted to improve the signal-to-noise ratio without degrading the resolution.
The polarized intensity P and the observed polarization angle $\Psi$ were obtained from the U and Q maps by applying the relation ${\rm P=\sqrt{Q^2+U^2}}$ and ${\rm \Psi=0.5\cdot \arctan(U/Q)}$.
The polarization maps were corrected for the positive bias introduced when combining U and Q images \citep[see Appendix B in][]{killeen}.
\subsection*{K-Band}
At 19\,GHz, the diffuse emission of CJ2242 is very faint because of its steep spectrum.
For instance, to detect the northern relic at 19\,GHz with a S/N=3, we would need to reach a sensitivity level of ${\rm\sigma}$=0.1\,mJy/beam in the SRT images.
Even with the SRT K-band multi-feed, this would require more than 100 hours of exposure time, too much to fit within the time allocated to the SMOG program.\\
Indeed, the K-band observations were aimed at characterizing the spectral behaviour of a few discrete sources embedded in the diffuse emission of the galaxy cluster CJ2242.
In this way we can have an accurate estimate of their flux densities at 6.6\,GHz and 
disentangle the discrete and the diffuse emission of CJ2242 (see \ref{sec:cmeas}).\\
We performed total intensity continuum observations, using a total bandwidth of 2\,GHz centred on 19\,GHz, at which the FWHM is $\sim 1^{\prime}$. 
The coordinates of the cross-scans observations and other details are reported in Table \ref{tab:cs}. \\
After the baseline subtraction from each cross scan we calibrated the flux density with 3C\,147
corrected for the atmosphere opacity and the gain variation with elevation.
We used sky-dip observations to infer the opacity $\tau$.
The opacity during our observations was $\tau \simeq$0.02.
Finally, we projected the data in a regular 2-dimensional grid with a cell-size of 15$^{\prime\prime}$/pixel and we stacked all the cross scans together to improve the signal-to-noise ratio.
We fitted the cross scans with a 2-dimensional Gaussian to derive the high frequency flux density.

\begin{figure*}\centering
\includegraphics[scale=1.2]{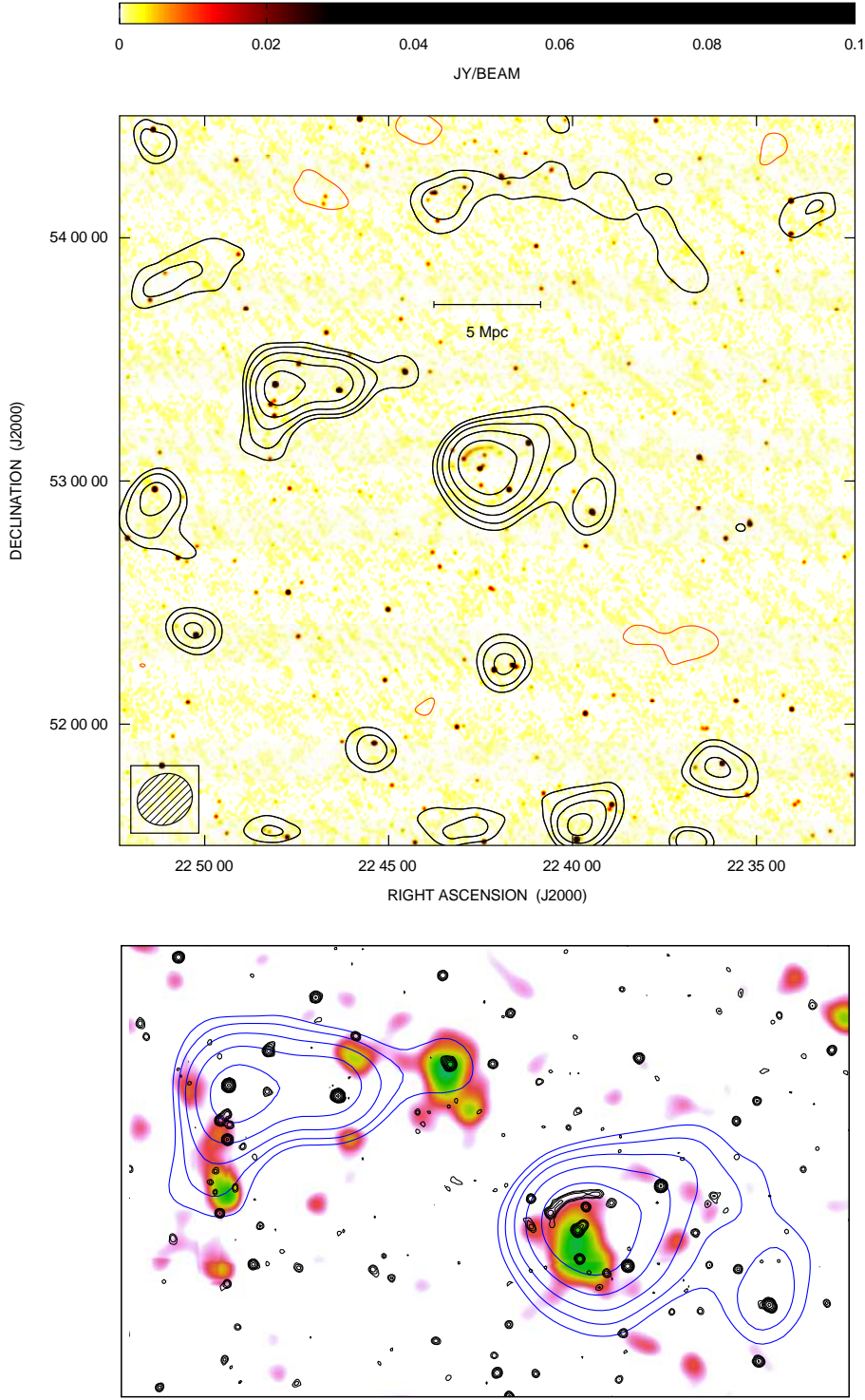}
 \caption{\textit{Top:} NVSS image overlaid with the SRT total intensity contours of the galaxy cluster CIZA J2242.8+5301 
 obtained with the L-band receiver in the frequency range 1.3-1.8\,GHz. 
 The field of view of the image is 3$^{\circ}$x3$^{\circ}$. 
 The FWHM beam is 14$^{\prime}$ and is shown in bottom-left corner.
 The noise level is 20 mJy/beam.
 Contours start at 3${\rm\sigma}$-level and increase by a factor of ${\rm\sqrt{2}}$ with negative contours (-3${\rm\sigma}$) drawn in orange.
 \textit{Bottom:} A zoomed version including the L-shaped structure (see text) and CJ2242 is shown. 
 The SRT contours in blue and the NVSS contours in black are overlaid on the X-ray image taken from the RASS in the 0.1-2.4 keV band.}
 \label{fig:lband}
\end{figure*}
\section{Total intensity results}
\subsection{L-Band}
\subsubsection{Image}
\label{sect:lband}
The results obtained from the L-band observations of CJ2242 are shown in Figure \ref{fig:lband} (top panel):
contours start at 3${\rm \sigma}$ with ${\rm \sigma}$=20\,mJy/beam and colours refer to the NVSS image of the cluster.
The central emission belongs to the galaxy cluster CJ2242. 
In the SRT image we note two additional features.
About one degree north of CJ2242 we see a diffuse arc-shaped structure which is likely due to a blending of discrete radio sources.
North-east of CJ2242, at $\sim$10\,Mpc from the cluster centre, we detect an extended ``L-shaped'' structure
which appears in the NVSS as a clustering of several point-like sources.
In the bottom panel of Figure \ref{fig:lband} we show a zoomed figure including this extended structure and CJ2242.
Colours refer to the X-ray image taken from the ROSAT All-Sky Survey (RASS, \citealt{rosat}) in the 0.1-2.4 keV band,
corrected for the background, divided by the exposure map, and smoothed with a Gaussian of ${\rm \sigma}$ =90$^{\prime \prime}$.
We overlaid the SRT contours in blue and those from NVSS in black.
The SRT L-shaped structure seems to connect a few spots of X-ray emission: 
(1) on the west side, the closest to CJ2242, hosts at its centre an NVSS source; (2) 
another source, which is seen at the southern tip of the extended L-shaped structure, is somewhat fainter and overlaps in projection
with several NVSS sources.\\
We wondered if the spatial coincidence between the radio and X-ray emission may possibly indicate the presence of one or more galaxy clusters, nearby to CJ2242.
The association with high hardness ratios of sources in the ROSAT Bright Source Catalogue \citep[BSC,][]{voges99} and in the Faint Source Catalogue (FSC),
coupled with the presence of an optical over-density or Sunyaev-Zel\textasciigrave dovich (SZ) signal, 
is a good indication of the presence of a cluster \citep[e.g.][]{ebeling,planckiv}.
A bright source with a high hardness ratio, associated with optical over-density, is the selection criterion of the CIZA catalogue itself \citep[][and references therein]{koc}. 
This criteria can be extended to faint sources \citep{ebeling13}.
The source closest to CJ2242 (labelled S1 in Sect. \ref{sect:s1}) is associated with the FSC source 1RXS J224504.3+532800, which has a hardness ratio of 0.93$\pm$0.09. 
This is also confirmed by the second RASS source catalogue \citep[2RXS,][]{boller} and the association with the source 2RXS J224454.9+532719 having a hardness ratio of 1.0$\pm$0.1.
See also Section \ref{sect:s1}.

\begin{figure*}\centering
 \includegraphics[width=1\textwidth]{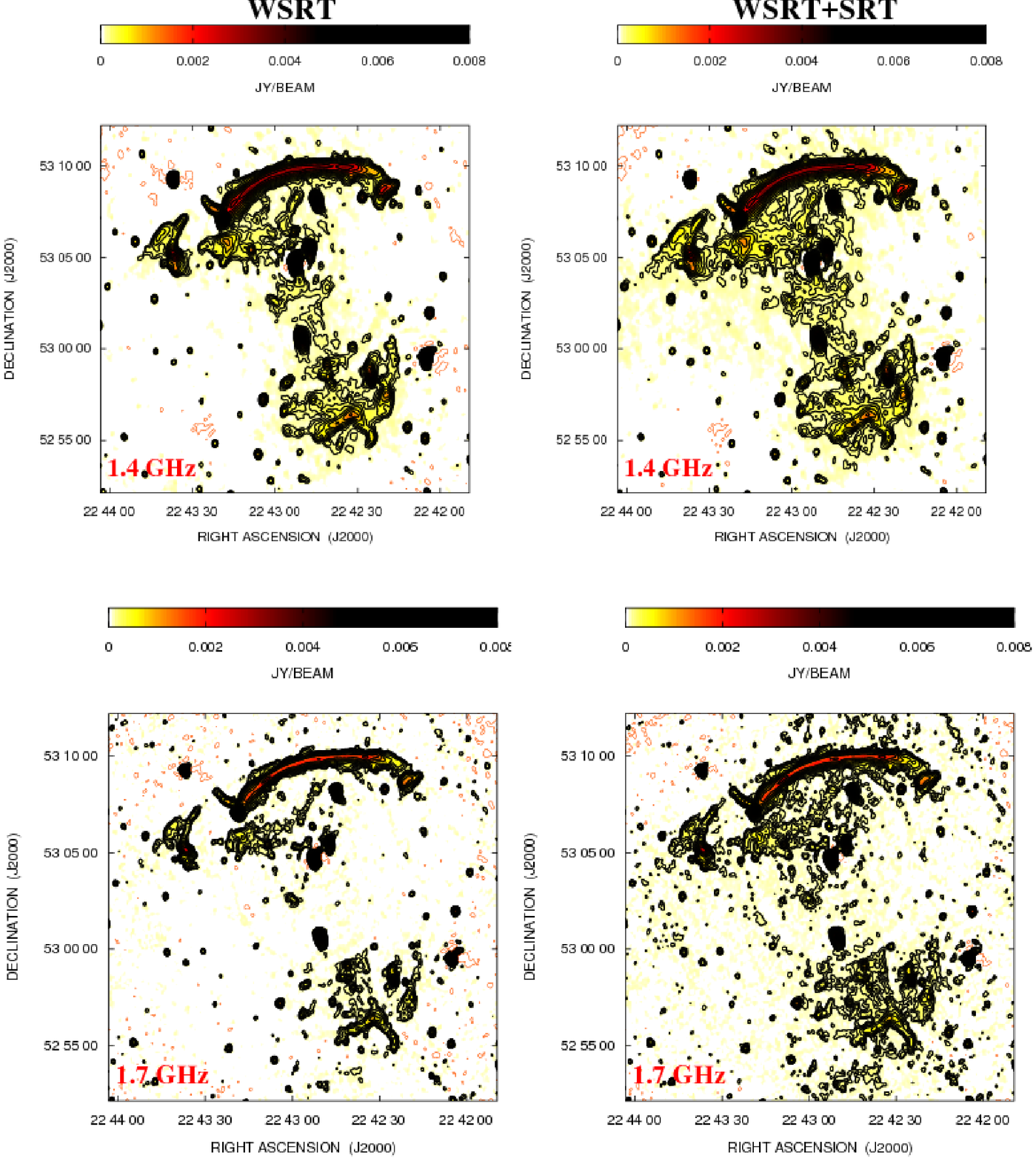}
 \caption{Results of the single-dish interferometer combination: on the left we show the WSRT maps at 1.4\,GHz (top) and 1.7\,GHz (bottom) taken from \citet{stroe2013}
 and on the right the WSRT and SRT combination maps.
Contours start at 3${\rm \sigma}$-level, where ${\rm \sigma}$=40\,${\rm\mu}$Jy/beam at 1.4\,GHz and ${\rm\sigma}$=30\,${\rm\mu}$Jy/beam at 1.7\,GHz, and increase by a factor of ${\rm\sqrt{2}}$. 
Negative contours (-3${\rm\sigma}$) are drawn in orange.
The beam sizes of the images are ${\rm 20.95^{\prime\prime}\times15.80^{\prime\prime}}$ at 1.4\,GHz and ${\rm15.98^{\prime\prime}\times13.10^{\prime\prime}}$ at 1.7\,GHz.}
 \label{fig:combo_sdi}
\end{figure*}
\subsubsection{Combination of single-dish and interferometer data}
We combined our single-dish map together with interferometric maps from \citet{stroe2013}, available online\footnote{http://vizier.cfa.harvard.edu/viz-bin/VizieR?-source=J/A+A/555/A110},
taken with the WSRT at 1.4 and 1.7\,GHz. Data were collected in the frequency ranges between 1.303$-$1.460\,GHz and 1.642$-$1.780\,GHz.
The beam sizes of the WSRT images are ${\rm 20.95^{\prime\prime}\times15.80^{\prime\prime}}$ and ${\rm 15.98^{\prime\prime}\times13.10^{\prime\prime}}$ at 1.4\,GHz and 1.7\,GHz respectively. 
Their sensitivity is ${\rm \sigma}$=40\,${\rm\mu}$Jy/beam at 1.4\,GHz and ${\rm\sigma}$=30\,$\mu$Jy/beam at 1.7\,GHz.\\
As already mentioned in the introduction, the angular extent of the diffuse emission in CJ2242 (radio halo and relics) exceeds 15$^{\prime}$.
The maximum structure that can be recovered by the WSRT at L-band, given the minimum baseline of ${\rm b_{min}}$=36\,m between the antennas of the array,
is about 16$^{\prime}$ for a source at the zenith.
As a result, a fraction of the flux density from the radio halo and the relics could have been missed by the WSRT, due to the lack of information in the inner portion of the ${\rm (u,v)}$-plane. 
The single-dish SRT L-band image does not suffer from this limitation, since structures as large as the angular scale of the SRT image are retained (3 degrees).\\
Therefore we combined the SRT with the WSRT images to reconstruct the correct large scale structure while preserving the angular resolution of the interferometric observations \citep[see e.g.][]{comb}.\\
We did this in the image plane using the SCUBE software package, in a similar way as the Astronomical Image Processing System (AIPS) 
task IMERG or the Miriad \citep{miriad} task IMMERGE.
In particular, we used the WSRT images which are already corrected for the primary beam attenuation. 
These images consist of a single pointing and are blanked outside a circular region of 55.3$^{\prime}$ and 45$^{\prime}$, at 1.4 and 1.7\,GHz, respectively.
We extracted two smaller sub-bands from the full SRT bandwidth (1.3$-$1.8\,GHz), centred at the exact central frequencies and with the same bandwidths as the WSRT data. 
We cropped the two SRT images produced from these sub-bands in the same way as the WSRT images to contain the same region of the sky.
After that, we transformed the single-dish and the interferometric images to combine them in Fourier space. 
The diameter of the SRT, 64-m, is larger than the minimum baseline of the WSRT, therefore in the Fourier plane there is a region of overlap in which both images share the same power spectral density. 
This annulus in the Fourier space is used to cross-check the calibration of the two images and to calculate a scaling factor that we then applied to the single-dish data. 
In order to account for the different resolution of the two instruments, we deconvolved the images, by dividing both of them by the Fourier transforms of the corresponding Gaussian beams,
before calculating the scaling factor.\\
Due to the superior signal-to-noise ratio the scaling factor was calculated at 1.7\,GHz.
The required adjustment resulted in a scaling-up factor for the SRT image of 1.23 that we also applied to the SRT 1.4\,GHz image. 
After this scaling the two power spectra were merged using a weighted sum of the Fourier transforms. 
For the single-dish observations, data weights are set to 0 for all wave-numbers larger than the outer ring of the annulus, while they are set to 1 in the inner portion of the Fourier plane. 
In the annulus the weights linearly vary from 0 to 1. 
The interferometric data are weighted in a similar way but with swapped values for the weights.\\
The combined Fourier spectrum was then tapered by multiplying by the transform of the interferometer beam. 
The combined image, obtained by the anti-transform, has the same angular resolution of the original WSRT image and includes the large scale structures detected by the SRT.\\

Results are shown in Figure \ref{fig:combo_sdi}: 
at left we present the WSRT maps at 1.4\,GHz (top) and 1.7\,GHz (bottom) and on the right we show the WSRT and SRT combined maps.
Contours start at 3${\rm \sigma}$-level.
Thanks to the combination we can recover the flux density associated with the diffuse sources of the galaxy cluster CJ2242, 
revealing a greater extension of the central radio halo, especially at 1.4\,GHz.
At 1.7\,GHz we do not see a significant enhancement of the radio halo emission.
This could be consistent with the fact that typically these sources have a steep spectrum (${\rm \alpha}$=1.3)
such that at higher frequencies the associated flux could be lower than the rms of the map.  
Assuming the same size at the two frequencies the mean surface brightness at 1.7\,GHz should be ${\rm \langle I_{1.7GHz} \rangle=\langle I_{1.4GHz}\rangle\times0.485}$, 
a factor that takes into account the different beam areas of the two instruments.
At 1.4\,GHz, ${\rm \langle I_{1.4GHz}\rangle \sim 3.8 \times \sigma_{1.4GHz}}$ that implies ${\rm \langle I_{1.7GHz} \rangle \sim0.074}$\,mJy/beam. 
Thus, in order to detect the radio halo at 1.7\,GHz, we would need a noise level of  ${\rm \sigma}$=25\,${\rm\mu}$Jy/beam, 
lower than the rms of the image.
\begin{figure}\centering
 \includegraphics[scale=0.5]{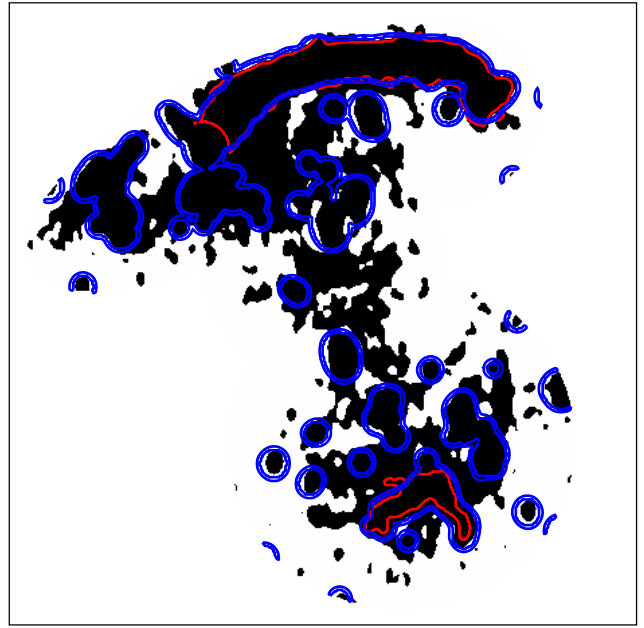}
 \caption{This image shows the area we considered to measure the flux of the diffuse sources hosted by CIZA J2242.8+5301: 
 red contours are drawn around the relic areas, while black colour corresponds to the radio halo emission. Blue contour indicates that strong sources that we do not 
 considered in the estimate of the halo mean surface brightness.}
 \label{fig:blank}
\end{figure}
\begin{figure}\centering
 \includegraphics[scale=0.4]{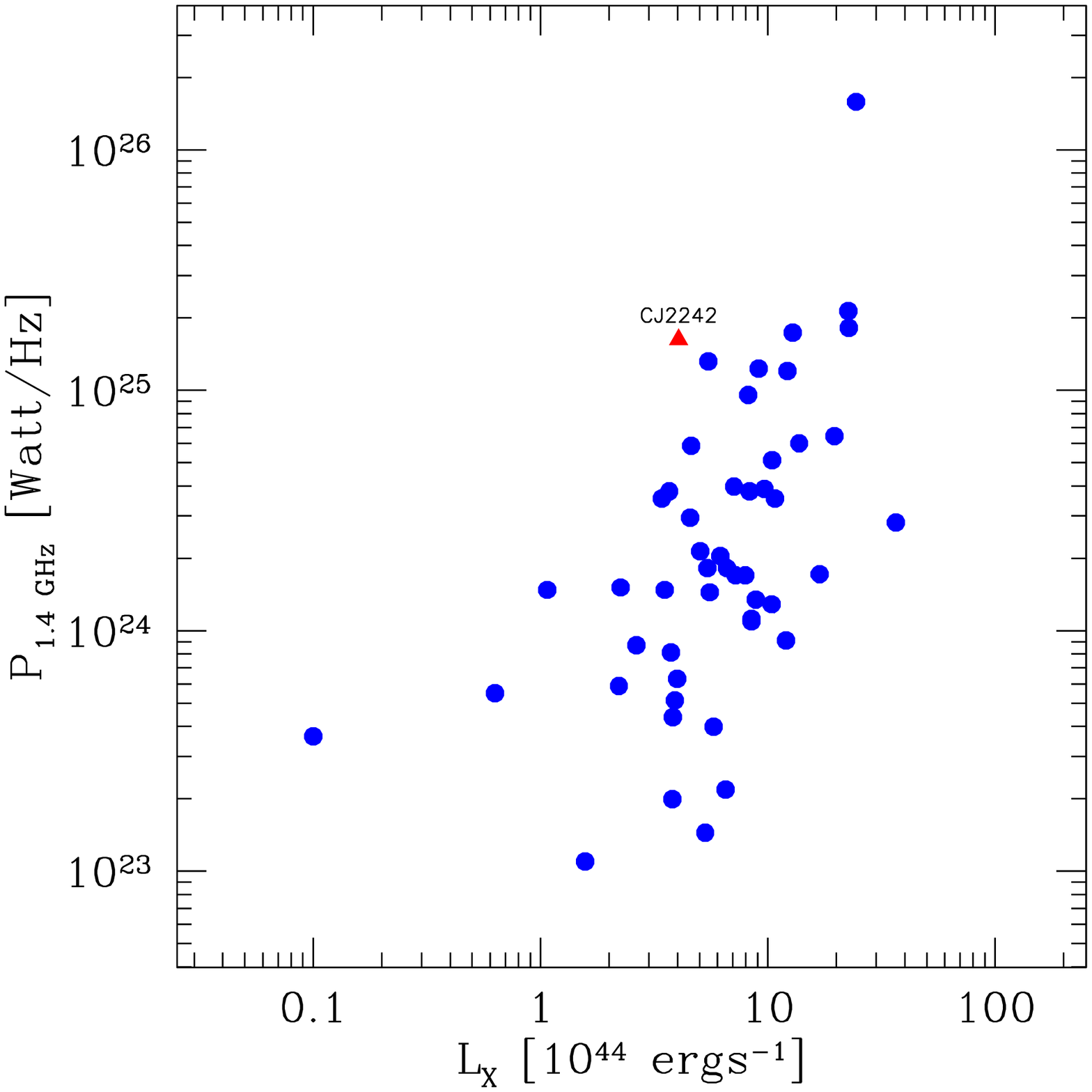}\\
  \includegraphics[scale=0.4]{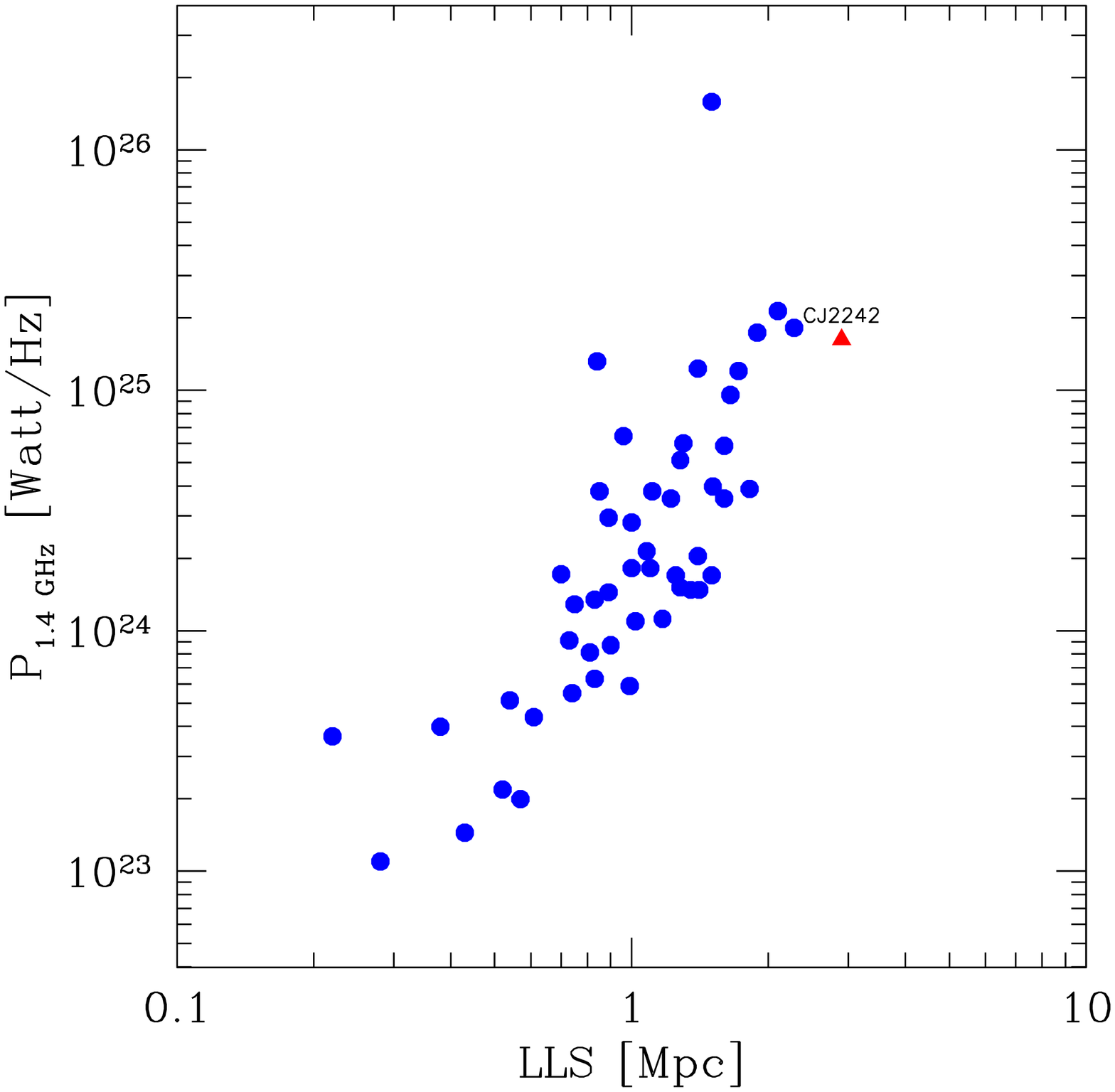}
 \caption{Radio halo power at 1.4\,GHz versus the cluster X-ray luminosity in the 0.1-2.4 keV band (top) and versus the largest linear size of the radio halo measured at 1.4\,GHz (bottom). 
 Full dots are observed clusters taken from the literature. We added to the compilation by \citet{feretti12} new radio haloes in merging galaxy clusters revealed through pointed 
 interferometric observations at $\simeq$1.4\,GHz \citep{fede12,van12,giovannini,martinez,shakouri,parekh}. A red triangle marks the cluster CJ2242.}
 \label{fig:stat}
\end{figure}
\subsubsection{Flux density measurements of diffuse sources}
We measured the flux density of the diffuse emission of the galaxy cluster CJ2242 at 1.4\,GHz using the combination of SRT and WSRT data.\\
As we do not observe radio halo emission in the WSRT 1.7\,GHz image of Figure \ref{fig:combo_sdi} at the 3${\rm\sigma}$-level we used this image
to blank the strong sources in the field of CJ2242 from the 1.4\,GHz images.
We ended up with an image where the discrete sources and the relics are blanked as shown by blue contours in Figure \ref{fig:blank}.\\
In this image black represents the emission above the 3${\rm\sigma}$-level of the combined map at 1.4\,GHz. 
We assumed that this is the extent of the radio halo and that the relics are located inside the red contours.
We defined the northern relic region by taking all those pixels inside the same area considered by \citet{stroe2016} in Figure A1 with a flux density greater than 5${\rm\sigma}$, 
excluding obvious discrete sources.
The southern relic emission is fainter than that of the northern relic.
Thus, to isolate the southern relic from the radio halo, we used a stronger limit in flux by drawing a 8${\rm\sigma}$-level contour around the relic.
We used the black area outside the blue contours (equivalent to $\sim$451 times the beam area) to evaluate a mean surface brightness of the radio halo. 
Then, we multiplied this value with the total area (equivalent to $\sim$869 times the beam area), 
as we assume that the halo extends even in the relic regions and that of the discrete sources.\\
We found that the radio halo hosted by CJ2242 has a total flux density of S$_{1.4GHz}^{SRT+WSRT}=(158.3\pm 9.6)$\,mJy. 
We noticed that if we measure with the same procedure the radio halo flux from the WSRT 1.4\,GHz image we find S$_{1.4GHz}^{WSRT}=(115\pm 7)$\,mJy.
In our estimates we included statistical and systematic (6$\%$ of the flux to include calibration uncertainties) errors.
The northern relic has a total flux density of S$_{1.4GHz}^{WSRT}=(121\pm 7)$\,mJy and S$_{1.4GHz}^{WSRT+SRT}=(126\pm 8)$\,mJy as calculated from the interferometric and combined maps respectively.
Here we have subtracted the contribution of the extended radio halo in order to take into account only the flux density of the northern relic.
Finally, we found that the southern relic has a flux density S$_{1.4GHz}^{WSRT+SRT}=(11.7\pm 0.7)$\,mJy, 
a value consistent, within errors, with that calculated from the interferometric map, S$_{1.4GHz}^{WSRT}=(11.0\pm 0.7)$\,mJy.\\
For the radio halo we find an enhancement of the flux density of about $\sim$ 38$\%$ when we combine single-dish and interferometric images, whereas
for the relics we obtain no significant differences. 
This demonstrates how the interferometer has poor sensitivity to extended and diffuse 
emission while it performs better with narrow 
structures such as the northern relic.\\

The statistical properties of radio haloes in galaxy cluster are important in order to understand the nature of these sources. 
Thanks to our measurements, we can now compare the properties of CJ2242 with other clusters that host radio haloes.
In Figure \ref{fig:stat}, we plot the 1.4\,GHz power of the radio halo versus the X-ray luminosity between 0.1$-$2.4 keV (top) and
the 1.4\,GHz power of the radio halo versus its largest linear size (bottom) measured at the same frequency.
Filled dots are observed clusters taken from the literature. We added to the compilation by \citet{feretti12} new radio haloes in merging galaxy clusters revealed through pointed 
interferometric observations at $\simeq$1.4\,GHz \citep{fede12,van12,giovannini,martinez,shakouri,parekh}. A red triangle marks the cluster CJ2242, one
of the largest radio haloes and one of the brightest.

\begin{figure*}
 \includegraphics[scale=0.6]{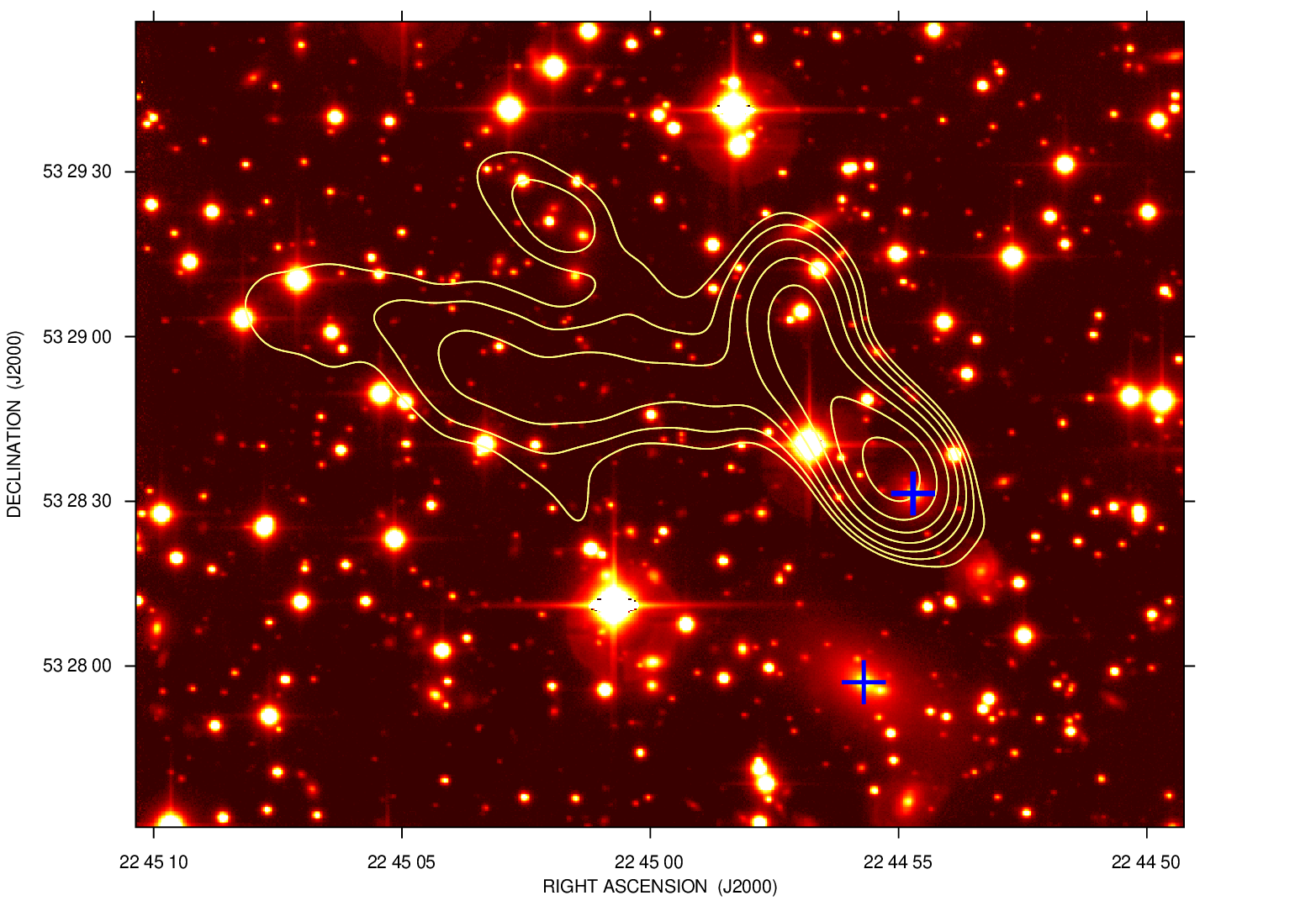}\\ 
  \vspace{0.75cm}
\includegraphics[scale=0.85]{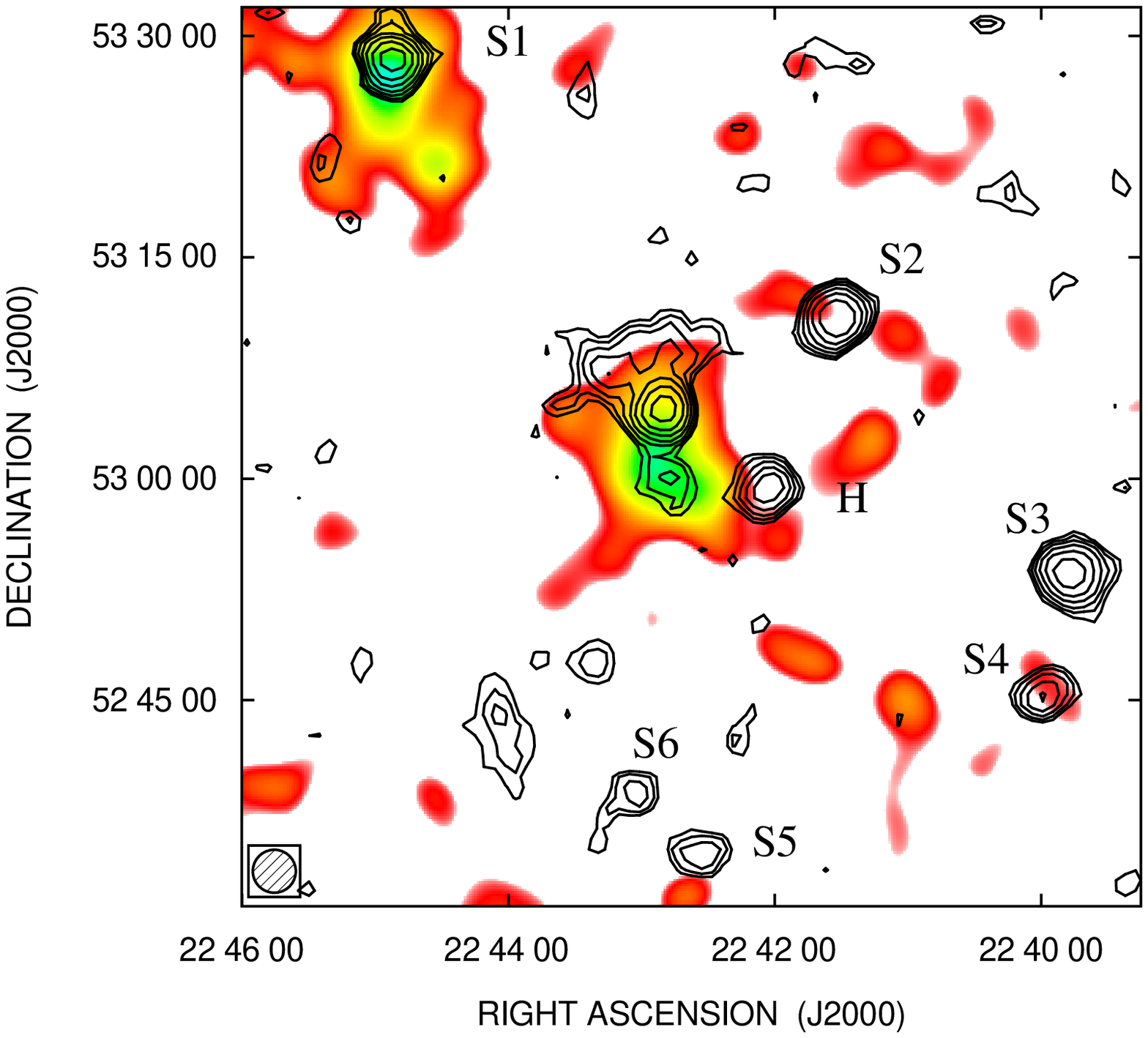}
 \caption{\textit{Top:} CFHT optical image of S1 with the two optical galaxies indicated by the crosses and contours from the 323 MHz GMRT image that start at 2\,mJy/beam and 
 increase by a factor 2.
 \textit{Bottom:} Shallow SRT total intensity contour image of the galaxy cluster CIZA J2242.8+5301 at 6.6\,GHz overlaid on the X-ray image taken from the RASS in the 0.1-2.4 keV band.
 The SRT image has been obtained from the spectral average of the data in the frequency range 6$-$7.2\,GHz. The field of view is 1$^{\circ}\times1^{\circ}$. 
 The FWHM beam is 2.9$^{\prime}$ and is shown in bottom-left corner.
 Contours start at 3${\rm\sigma}$-level, with ${\rm\sigma}$=1\,mJy/beam, and increase by a factor of ${\rm\sqrt{2}}$.
 }
 \label{fig:i_shallow}
\end{figure*}
\subsection{Shallow C-band}
\subsubsection{Image}
\label{sect:s1}
In Figure \ref{fig:i_shallow} (bottom) we show the resulting total intensity contour map in the frequency range between 6 and 7.2\,GHz over a region of 1$^{\circ}$x1$^{\circ}$.
We reached a noise level of ${\rm\sigma}$=1 mJy/beam with a beam size of $2.9^{\prime}\times2.9^{\prime}$.
The radio contours at 6.6\,GHz are overlaid on the same X-ray image as Figure \ref{fig:lband}, taken from the RASS in the 0.1-2.4 keV band.
The central cluster is bright both in the X-ray and radio band and we detected the diffuse radio emission associated with CJ2242.\\
Among the external sources, marked with a letter in the image, the most interesting is S1. 
As already pointed out in Section \ref{sect:lband}, the X-ray emission has the right hardness ratio to be classified as a galaxy cluster candidate.
We show in Figure \ref{fig:i_shallow} (top), contours of the 323 MHz GMRT image of \citet{stroe2013}.
These contours are drawn starting from 2\,mJy/beam and increase by a factor of 2.
Thanks to the higher resolution of the GMRT we can appreciate the head-tail morphology of the source which is further indication of the presence of a dense external medium 
similar to what is typically found in galaxy clusters.
The radio contours are overlaid on an optical Canada-France-Hawaii Telescope (CFHT) image\footnote{http://www.cadc-ccda.hia-iha.nrc-cnrc.gc.ca/en/search/},
where two blue crosses indicate the position of two optical galaxies:
2MASX J22445464+5328318, at less than 0.2$^{\prime}$ from S1, and 2MASX J22445565+5327578, at less than 0.7$^{\prime}$.
The galaxies have a K-band magnitude of M=(13.5$\pm$0.1) mag and M=(12.67$\pm$0.06) mag, respectively, already corrected for extinction \citep[see Appendix in][]{schlafly}.
By using the K-band magnitude-redshift relations of \citet{inskip} we found that these galaxies have a redshift z$\sim$0.1-0.2, similar to CJ2242.
We obtained a similar result ($z\sim0.15-0.2$) when considering the K-band
magnitude-redshift relation for brightest cluster galaxies \citep[see Fig.5 in][]{stott}.

Even if at the moment there are only two optical galaxies identified in the vicinity of S1, this fact, together with its X-ray properties and the fact that S1 is an head-tail radio source, 
corroborates the hypothesis of S1 as a cluster candidate.

\begin{table}\centering
\caption{Flux measurements of the sources marked in Figure \ref{fig:i_shallow} obtained from the NVSS 1.4\,GHz \citep{condon} and the SRT 6.6\,GHz images (this work).}
  \label{tab:source}
\begin{tabular}{l l l l l}
  \hline 
   Label & Name & ${\rm S_{1.4GHz}^{mJy}}$ & ${\rm S_{6.6GHz}^{mJy}}$ & ${\rm\alpha}$\\
  \hline 
  S1& NVSS J224455+532840   & 133$\pm$8    & 28$\pm$2 & 1.00$\pm$0.09\\   
  S2& 87GB 223927.9+525526  & 138$\pm$8    & 26$\pm$2 & 1.08$\pm$0.09\\   
  S3& 4C +52.50             & 182$\pm$11   & 26$\pm$2 & 1.3$\pm$0.1\\     
  S4& NVSS J224001+524538   & 35$\pm$2     & 9$\pm$1  & 0.9$\pm$0.1\\     
  S5& NVSS J224234+523506   & 32$\pm$2     & 6$\pm$1  & 1.1$\pm$0.2 \\    
  S6& NVSS J224258+523912   & 11$\pm$1     & 5$\pm$1  & 0.5$\pm$0.2 \\    
  H & NVSS J224205+525931   & 99.6$\pm$0.6 & 14$\pm$2 & 1.3$\pm$0.2\\     
  \hline
\end{tabular}
\end{table}
\subsubsection{Flux density measurements of discrete sources}
In Table \ref{tab:source} we give the flux density of the sources identified in the field and marked with a letter in Figure \ref{fig:i_shallow}, 
measured from the NVSS 1.4\,GHz and the SRT 6.6\,GHz images.
We also calculated the spectral index ${\rm\alpha}$ of the sources assuming for the flux density ${\rm S_{\nu}}$ a power law behaviour with the frequency ${\rm\nu}$,
\begin{equation}
 S_{\nu}=S_0 ~ \bigg( \frac{\nu}{\nu_0} \bigg)^{-\alpha} .
 \label{eq:flux}
\end{equation}
For source S1 we found $\alpha=1.00\pm0.09$ which is typical for head-tail sources as pointed out in the previous Section.

\begin{figure*}\centering
 \includegraphics[scale=0.7]{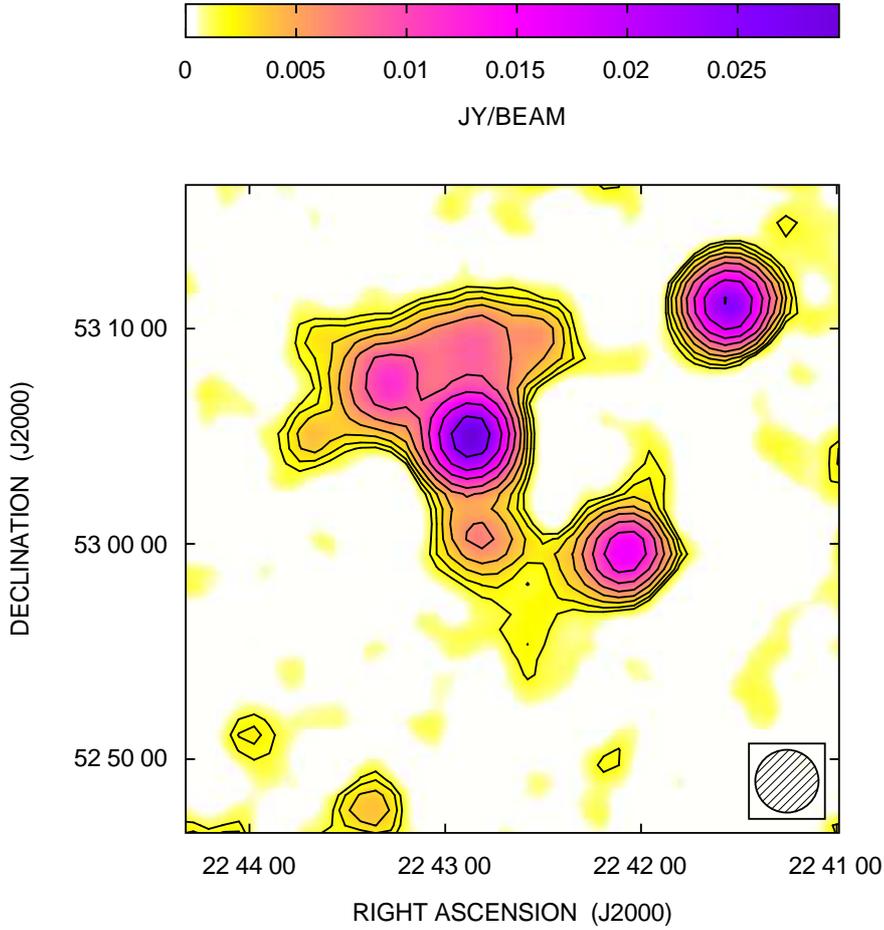}
 \caption{Deep SRT total intensity image of the galaxy cluster CIZA J2242.8+5301 obtained from the spectral average over the frequency range 6$-$7.2\,GHz. 
 The field of view of the image is 30$^{\prime}\times30^{\prime}$. 
 The FWHM beam is 2.9$^{\prime}$ and is shown in bottom-right corner.
 The noise level is 0.5 mJy/beam.
 Contours start at 3${\rm\sigma}$-level and increase by a factor of ${\rm\sqrt{2}}$.}
 \label{fig:i_deep}
\end{figure*}
\subsection{Deep C-Band}
\subsubsection{Image}
In Figure \ref{fig:i_deep} we show the total intensity map resulting from all the observations at 6.6\,GHz in a field-of-view of 30$^{\prime}\times30^{\prime}$ centred on the cluster centre.
In this image we reached a noise level of $\sigma=0.5$\,mJy/beam.
Here the two relics are clearly detected.

\begin{table*}\centering
  \caption{Flux density measurements obtained from the following maps:
  GMRT 323 MHz \citep{stroe2013}, WENSS 325 MHz \citep{wenss}, GMRT 608 MHz \citep{stroe2013}, NVSS 1400 MHz \citep{condon}, WSRT 1714 MHz \citep{stroe2013}, WSRT 2272 MHz \citep{stroe2013},
  VLA 4835 MHz and 4885 MHz \citep{van10}, SRT 19\,GHz (this work). We report the expected flux density of each source at 6.6\,GHz in the last column.}
  \label{tab:meas}
\begin{tabular}{l l l l l l l l l l l}
  \hline 
  {\bf Source} & {\bf ${\rm S_{323MHz}^{mJy}}$} & {\bf ${\rm S_{325MHz}^{mJy}}$} & {\bf ${\rm S_{608MHz}^{mJy}}$} & {\bf ${\rm S_{1.4GHz}^{mJy}}$} & {\bf ${\rm S_{1.7GHz}^{mJy}}$} & {\bf ${\rm S_{2.2GHz}^{mJy}}$} & {\bf ${\rm S_{4.835GHz}^{mJy}}$} & {\bf ${\rm S_{4.885GHz}^{mJy}}$} & {\bf ${\rm S_{19GHz}^{mJy}}$} & {\bf [${\rm S_{6.6GHz}^{mJy}}$]$^{\rm exp}$}\\ 
  \hline
  A & 56$\pm$6 &  53$\pm$5 & - & 17$\pm$2 & 19$\pm$2 & 17$\pm$2 & 6.1$\pm$0.6&6.0$\pm$0.6 & 3$\pm$1 & 4.8 \\
  B & 64$\pm$6 & 51$\pm$5 & 43$\pm$4 & 16$\pm$2 & 18$\pm$2 & 15$\pm$2 & 5.8$\pm$0.6 & 5.7$\pm$0.6 & 3$\pm$1 & 4.6\\
  C & 80$\pm$8 & - & 51$\pm$5 & - & 26$\pm$3 & 22$\pm$2 & 8.9$\pm$0.9 & 8.2$\pm$0.8 & 4$\pm$1 & 6.9\\
  D & 54$\pm$5 & -& 35$\pm$3 & -& 17$\pm$2 & 12$\pm$1 & 5.4$\pm$0.5 & 5.2$\pm$0.5 & 6$\pm$3 & 4.0 \\
  E & 357$\pm$36 & - & 248$\pm$25 & - & 106$\pm$11 & 82$\pm$8 & 39$\pm$4 & 39$\pm$4 & 11$\pm$1 & 31.0\\
  F & - & 44$\pm$5 & - & 13$\pm$2 & 11$\pm$1 & 8.2$\pm$0.8 & 2.4$\pm$0.2 & 2.8$\pm$0.3 & - & 1.4\\
  G & 50$\pm$5 & 46$\pm$5 & 38$\pm$4 & 42$\pm$4 & 19$\pm$2 & 18$\pm$2 & 14$\pm$1 & 6.8$\pm$0.7 & 3$\pm$2 & 5.4\\
  H & 102$\pm$10 & - & - & 102$\pm$10 & 80$\pm$8 & 62$\pm$6 & 24$\pm$2 & 24$\pm$2 & 5$\pm$2 & 17.4\\
  I & 20$\pm$2 & - & 13$\pm$1 & 9$\pm$1 & 6.6$\pm$0.7 & 5.2$\pm$0.5 & 2.1$\pm$0.2 & 2.2$\pm$0.2 & - & 1.4\\
    \hline
\end{tabular}
\end{table*}
\begin{figure*}\centering
  \includegraphics[scale=1]{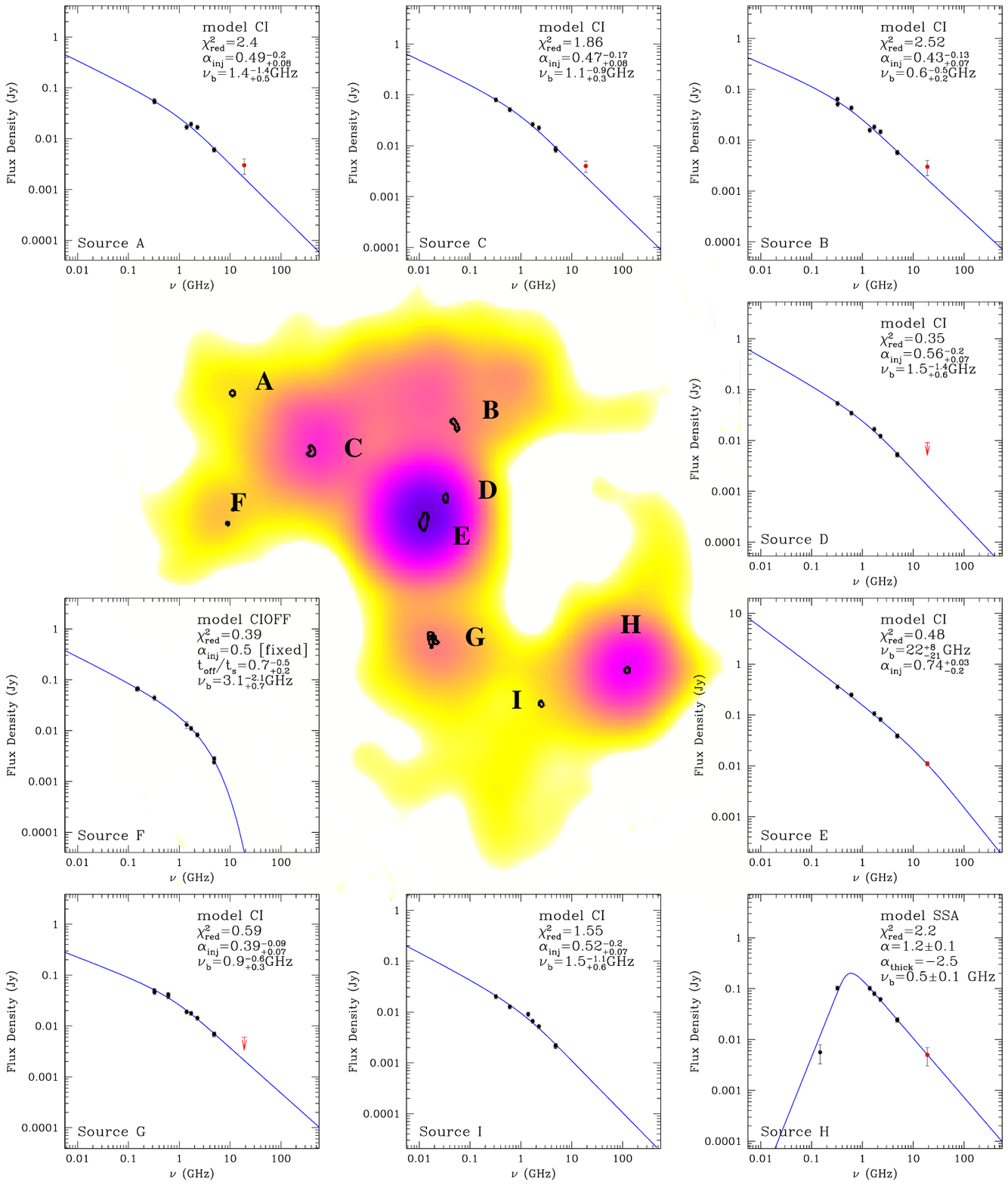}
  \caption{The image is the deep 6.6\,GHz SRT total intensity map of the galaxy cluster CIZA J2242.8+5301 with VLA contours at 5 ${\rm\sigma}$-level (${\rm\sigma}$=0.03 mJy/beam).
  Each plot around the image refers to the spectrum of the discrete source specified in the bottom left corner. 
  Black points are the flux density measurements from GMRT, WSRT, WENSS, NVSS and VLA images, while red points indicate the 19\,GHz SRT measurements (see text for details). 
  The continuous blue line is the result of the fit whose parameters are reported in the top right corner of each panel.}
  \label{fig:sources}
\end{figure*}
\begin{table}\centering
  \caption{Results of the spectral modelling and source ages.}
  \label{tab:ages}
\begin{tabular}{l l l l l l}
  \hline 
  Source & Model & ${\rm \alpha_{inj}}$ & ${\rm\nu_b}$\,[GHz] &  ${\rm B_{min}\,[\mu G]}$ & ${\rm t_s}$\,[Myr]\\ 
  \hline
  A & CI & 0.5$\pm$0.1 & 1.4$\pm$1.0 & - & -  \\
  B & CI & 0.4$\pm$0.1 & 0.6$\pm$0.4 & 8.1 & 76$\pm$36 \\
  C & CI & 0.5$\pm$0.1 & 1.1$\pm$0.6 & 6.9 & 70$\pm$24 \\
  D & CI & 0.6$\pm$0.1 & 1.5$\pm$1.0 & 8.4 & 46$\pm$22\\
  E & CI & 0.7$\pm$0.1 & 22$\pm$14 & 10.2 & 45$\pm$16 \\
  F & CI$_{\rm OFF}$ & 0.5 & 3$\pm$1 & 1.4 & 149$\pm$60  \\
  G & CI & 0.4$\pm$0.1 & 0.9$\pm$0.5 & 3.6 & 166$\pm$58 \\
  H & SSA+PL & 0.7$\pm$0.1 & <0.54 & - & - \\
  I & CI & 0.5$\pm$0.1 & 1.5$\pm$0.6 & - & - \\
    \hline
\end{tabular}
\end{table}
\subsubsection{Discrete sources modelling}
\label{sec:cmeas}
We characterize the spectral behaviour of the discrete sources, at less than 15$^{\prime}$ from the cluster centre, by using
the following multi-frequency high resolution images of the galaxies:
\begin{enumerate}
 \item from the NRAO archive we retrieved observations from the program AV312 presented by \citet{van10}, which used the VLA in C configuration at 4835 MHz and 4885 MHz. We calibrated data 
following standard procedures using the AIPS software package;
 \item we used GMRT 323 MHz, 608 MHz and WSRT 1714 MHz, 2272 MHz images from \citet{stroe2013} available online\footnote{http://vizier.cfa.harvard.edu/viz-bin/VizieR?-source=J/A+A/555/A110};
 \item we included in our sample the Westerbork Northern Sky Survey (WENSS) at 325 MHz \citep{wenss} and the NVSS at 1400 MHz \citep{condon} maps;
 \item we also added the SRT 19\,GHz observations  reported in  Table \ref{tab:cs}.
 \end{enumerate}
We considered the nine sources marked in Figure \ref{fig:sources} whose flux is mixed with the diffuse sources at the SRT resolution.
In this image colours represent the SRT total intensity image at 6.6\,GHz, and contours are from the VLA 4.8\,GHz contours at 5 ${\rm\sigma}$-level.   
We measured the flux density of each point-like source using a two-dimensional Gaussian fit, while for the extended sources we integrated the radio brightness down to the first isophote 
above the noise level. 

In Table \ref{tab:meas} we present the values obtained including a systematic error of 10$\%$ of the source flux, to take into account the different calibration scales of the images.
Figure \ref{fig:sources} shows the resulting spectra: in each panel the corresponding source is indicated in the bottom left corner
with red dots indicating SRT 19\,GHz measurements listed in Table \ref{tab:meas} together with other measurements shown by black dots.
Spectra are modelled by using the software SYNAGE \citep{murgia99}.
The continuous blue line is the result of the fit using parameters as reported in the top right corner of each panel.\\

The spectra of A, B, C, D, E, G, and I are well described by a continuous injection (CI) model \citep{ci}, 
where the sources are continuously and constantly replenished with new relativistic electrons by the AGN and the 
power-law spectrum breaks due to energy losses caused by the synchrotron radiation itself and inverse Compton scattering with the Cosmic Microwave Background (CMB) photons.
The CI model includes three free parameters: the injection spectral index, ${\rm \alpha_{inj}}$, that characterizes the spectrum below the break frequency, ${\rm \nu_b}$, and the normalization.
Above ${\rm \nu_b}$, the high-frequency spectral index is:
\begin{equation}
 \alpha_{h}=\alpha_{inj}+0.5
 \label{eq:alpha}
\end{equation}
From these parameters we can evaluate the synchrotron age of the source ${\rm t_s}$, i.e. the time elapsed since the start of the injection: 
\begin{equation}
 t_s=1590 \frac{B^{0.5}}{(B^2+B_{IC}^2)[(1+z)\nu_b]^{0.5}} \, {\rm Myr}
 \label{eq:ts}
\end{equation}
where B is the magnetic field in ${\rm\mu}$G of the source and ${\rm B_{IC}}$ is the equivalent magnetic field due to the inverse Compton of the CMB 
that depends on redshift ${\rm B_{IC}=3.25(1+z)^2 \,\mu}$G.
The break frequency is in\,GHz.
We calculated the synchrotron age by adopting the equipartition value ${\rm B_{min}}$ for the magnetic field strength \citep[see Eq. A.12 in][]{murgia12}.
Since this value depends on the source volume we derived ${\rm B_{min}}$ and hence the age only for the extended sources B, C, D, E and G.
The results of the spectral modelling and the source ages are listed in Table \ref{tab:ages}.
\paragraph*{Source A:}
It is point-like, the spectrum is fitted by a CI-model with ${\rm \alpha_{inj}=0.5}$, and ${\rm \nu_{b}=1.4}$\,GHz. The observed spectrum includes SRT measurements at 19\,GHz.
\paragraph*{Source B:}
The spectrum for this source is described by a CI-model, with ${\rm \alpha_{inj}=0.43}$ and ${\rm \nu_{b}=0.6}$\,GHz. It is a small double of 164\,kpc in size.
Assuming a cylindrical volume we calculated an equipartition magnetic field of ${\rm B_{min}\approx 8.1\,\mu G}$,
and we deduce an age of ${\rm t_s}$=76\,Myr.
\paragraph*{Source C:}
This is a narrow-angle-tail source, seen in projection over the northern relic. 
The tail, which has a linear size of 146\,kpc, extends toward the north-east.
We calculate for this source an age of ${\rm t_s}$=70\,Myr.
\paragraph*{Source D:}
This source is seen in projection close to the cluster centre, and is slightly extended at the resolution of the VLA 4.8\,GHz image.
The spectrum is well fit by a CI-model. We estimate an age of 46\,Myr.
\paragraph*{Source E:}
This source is located in projection close to the cluster centre. Its morphology shows two bright lobes and two faint tails extending in the N-S direction.
The spectrum is fitted by a CI-model with ${\rm \nu_{b}=}$22\,GHz and has an estimated age of ${\rm t_s}$=45\,Myr.
\paragraph*{Source F:}
This source is located at the Eastern tip of the northern relic. 
For this source we added the flux density measurements we derived from the 150 MHz TIFR GMRT Sky Survey 
(TGSS\footnote{http://tgssadr.strw.leidenuniv.nl/doku.php}) \citep{intema} and the WSRT 153 MHz \citep{stroe2013} images:
${\rm S_{150MHz}=(65\pm6)}$\,mJy and ${\rm S_{153MHz}=(67\pm2)}$\,mJy.
The radio spectrum shows a sharp high frequency exponential cut-off that cannot be explained 
by the smooth steepening of the CI-model. 
This source seems to be an example of a dying source where the central black hole of the galaxy has 
stopped its activity \citep{murgia11}. 
Following these authors, we fitted the radio spectrum using the CI$_{\rm OFF}$ model. 
This model assumes that the CI phase is followed by a remnant (or dying) phase during which the radio jets are switched off.
In the absence of the injection of new electrons the sources spectrum develops an exponential high frequency cut-off.
By fitting the CI$_{\rm OFF}$ model we derived the break frequency which gives the total source age, and t$_{\rm OFF}$/t$_{\rm s}$, which gives the relative duration of 
the dying phase.
We found a total source age of 149\,Myr. The source has been in the active phase for 45\,Myr and in the dying phase for 104\,Myr.
The morphology is relaxed, as expected for dying sources, but a weak point-like core is present.
This weak core seems however to be disconnected from the fading lobes as no jets are visible.\\
Source F could be a potential source of seed electrons for the northern relic \citep[e.g.,][]{bonafede14,van17}.
By looking at the top-right panel of Figure \ref{fig:combo_sdi} we see that the source is close to an arc-like feature 
which appears to be a secondary shock or a possible extension of the northern relic, although a discontinuity between these two structures is present.
\paragraph*{Source G:}
This is a narrow-angle tail with  a linear size of 174\,kpc.
The spectrum is well described by a CI model, with ${\rm \nu_b}\approx$0.9\,GHz.
The source age is 166\,Myr from which we deduced an advancing speed for the galaxy of 1000$\pm$400 $\cdot {\rm \frac{1}{\sin(i)}}\,$km/s,
where $i$ is the inclination of the source with respect to the line-of-sight.
\paragraph*{Source H:}
For this source we added the flux density measurement derived from the 150 MHz TGSS image: ${\rm S_{150MHz}=(5.6\pm2.3)}$\,mJy.
This source exhibits a spectral turn-over at low frequencies, that could be due to synchrotron self-absorption (SSA).
In the optically thick part of the spectrum we fixed the spectral index to $\alpha_{\rm thick}$=-2.5, while we modelled the optically thin regime using
a power-law. 
The source is unresolved and could be intrinsically a compact steep spectrum source where the break frequency is below the SSA peak: $\nu_{\rm b}$<<$\nu_{\rm SSA}$. For this
reason the observed spectral index in the optically thin part of the spectrum is interpreted as $\alpha=\alpha_{\rm inj}+0.5$.
\paragraph*{Source I:}
The spectrum of this source is well fit by a CI model with the break at 1.5\,GHz and $\alpha_{\rm inj}\approx$0.5. The source is point-like.\\

From the fitted spectra we derived the flux density of the sources at ${\rm\nu}$= 6.6\,GHz and report these in the last column of Table \ref{tab:meas}.

\begin{figure*}\centering
 \includegraphics[scale=0.7]{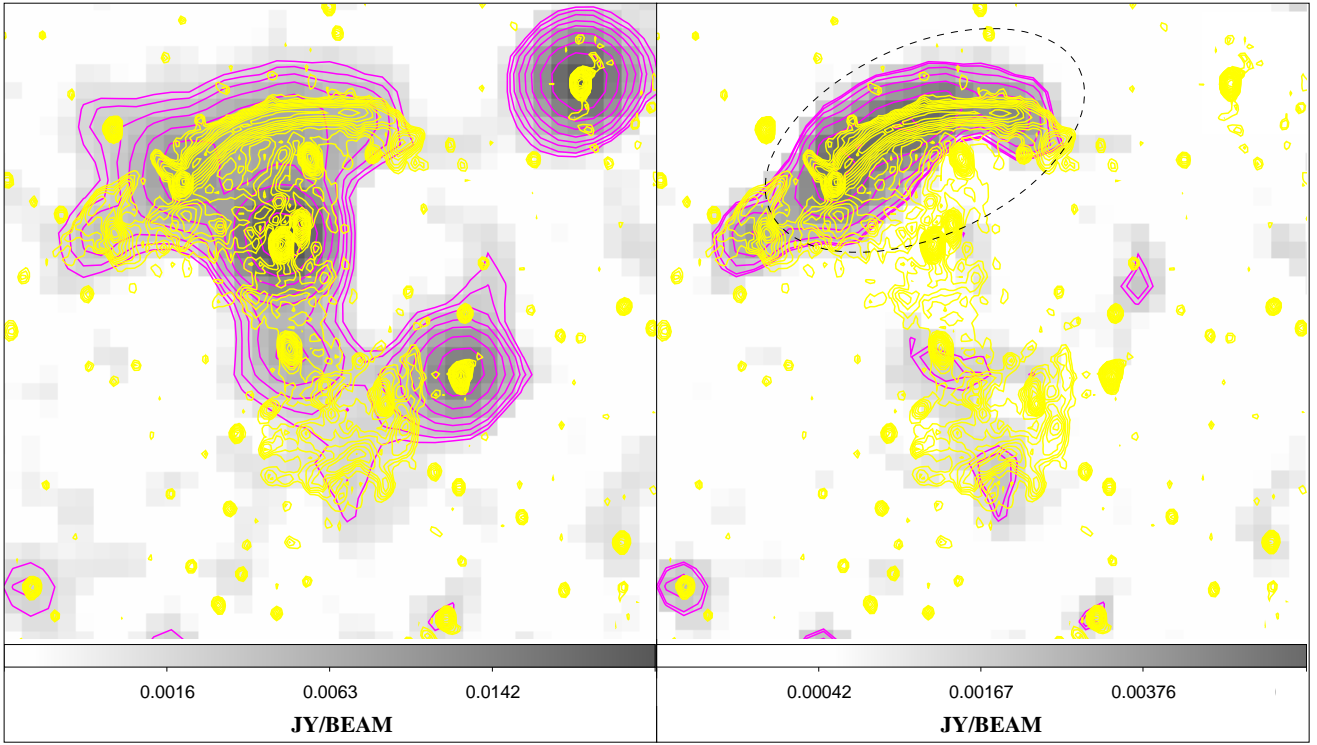}
 \caption{\textit{Left:} The original SRT image at 6$-$7.2\,GHz (colour-scale + magenta contours) overlaid with the higher resolution WSRT+SRT 1.4\,GHz contours (yellow).
 \textit{Right:} The residual 6$-$7.2\,GHz image of the diffuse emission after the subtraction of modelled discrete sources (see text for details). 
 Again the WSRT+SRT 1.4\,GHz contours are overlaid in yellow.}
\label{fig:sub}
\end{figure*}
\subsubsection{Flux density measurements of diffuse sources}
As mentioned in the introduction, single-dish telescopes can be used to accurately infer the size and flux density of diffuse sources.
They are not affected by the missing zero spacing problem that plagues radio interferometers, especially at high frequency
where the primary beam is usually smaller than the cluster size in the local Universe.
Unfortunately, the typical beam of single-dish telescopes is far larger than the beam synthesised by the interferometer, so it is difficult to distinguish between diffuse and discrete sources.\\
Thanks to the spectral modelling of the discrete sources from the previous section
we have a firm prediction of their flux densities at 6.6\,GHz (see last column of Table \ref{tab:meas}).\\
We modelled these sources with Dirac delta functions normalized to the expected fluxes.
We then convolved these functions with the SRT beam and 
finally we subtracted the resulting image from our SRT 6.6\,GHz image.\\
Figure \ref{fig:sub} shows in grey-scale and magenta contours the total intensity deep image on the right and, on the left, 
the image with the sources subtracted, where we expect to observe the contribution of diffuse emission only.
The noise is ${\rm\sigma}$= 0.5 mJy/beam. 
Both images show yellow contours of the combined image at 1.4\,GHz presented in Figure \ref{fig:combo_sdi}.
At the 3${\rm\sigma}$-level the northern emission seems to extend beyond the strong galaxy C covering a total extension of $\sim$2.9 Mpc and a width of $\sim$800\,kpc.
However, the extra-component in the north-east direction could be either due to a real extension of the northern relic or to the residual flux from source F (see notes on Sect. \ref{sec:cmeas}).
For these reasons, and to be consistent with the other measurements \citep[see Fig. A1 of][]{stroe2016}, we evaluated the northern relic flux density inside the ellipse drawn with a dashed line in Figure \ref{fig:sub}.
At the 5${\rm\sigma}$-level \citep[a cut consistent with that of][]{maja}, we found ${\rm S_{6.6GHz}=(17.1\pm1.2)\,mJy}$.\\
From Figure \ref{fig:sub} we can also give an estimate of the flux of the southern relic at 6.6\,GHz.
We measured a flux density of ${\rm S_{6.6\,GHz}=(0.6\pm0.3)\,mJy}$ at the 3${\rm\sigma}$-level.

\begin{table}\centering
  \caption{Flux density measurements of the northern relic of CIZA J2242.8+5301.}
  \label{tab:rn_meas}
\begin{tabular}{l l l}
    \hline
    ${\rm\nu}$ [GHz] & ${\rm S_{\nu}}$ [mJy] & Reference \\
    \hline
      0.15 & 780.4$\pm$80 & \citet{stroe2016} \\
      0.325 & 315.7$\pm$32.4 & \citet{stroe2016} \\
      0.327 & 446$\pm$21 & This work, WENSS image \\
      0.61 & 222.3$\pm$22.4 & \citet{stroe2016} \\
      1.2  & 125.7$\pm$12.6 & \citet{stroe2016} \\
      1.4  & 126$\pm$8 & This work, SRT image \\
      1.7  & 91.2$\pm$9.2 & \citet{stroe2016} \\
      2.25 & 61$\pm$3.6 & \citet{stroe2016} \\
      2.3 & 54.3$\pm$5.6 & \citet{stroe2016} \\
      4.85 & 32$\pm$8 &  \citet{maja} \\
      6.6 & 17.1$\pm$1.2 & This work, SRT image\\
      8.35 & 17$\pm$5 & \citet{maja} \\
    \hline
\end{tabular}
\end{table}
\begin{figure}
 \includegraphics[scale=0.4]{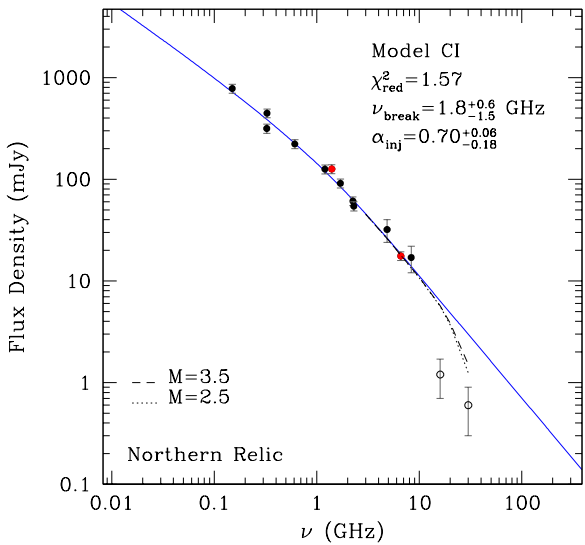}
  \caption{Total flux measurements of the northern relic at different frequencies from Table \ref{tab:rn_meas}. 
  Red points are the new measurements from this paper obtained with the SRT observations.
  The blue line represents the continuous injection model fit whose parameters are shown in the top-right corner of the image. 
  Dot and a dashed lines have been drawn to show the flux density decrement due to the SZ effect as predicted for the northern relic of CJ2242 by \citet{basu} assuming M=2.5 and M=3.5 
  respectively.}
  \label{fig:plot}
\end{figure}
\begin{figure}
 \includegraphics[scale=0.4]{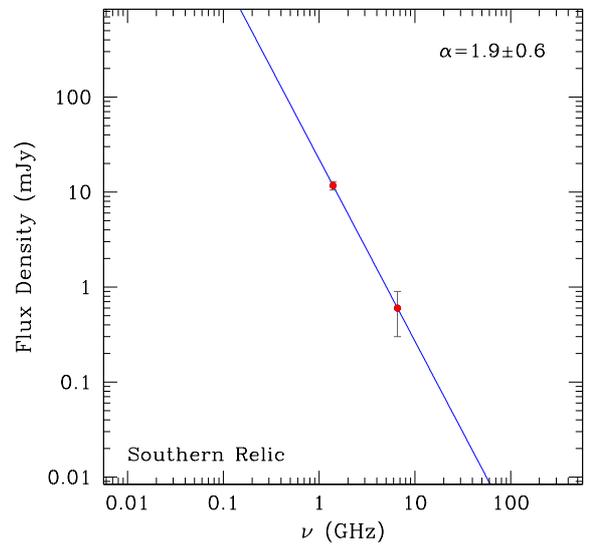}
  \caption{Total flux measurements of the southern relic obtained with the SRT observations.}
  \label{fig:sr}
\end{figure}
\subsection{Northern and southern relics}
\label{sect:nr_spix}
Our results regarding the flux of the northern relic are reported in Table \ref{tab:rn_meas}
together with flux density measurements at different frequencies taken from literature up to 8.4\,GHz. 
We included the value we inferred from the WENSS 327 MHz image after the subtraction of sources A, B and C, with the flux densities expected from our models at 327 MHz.

In Figure \ref{fig:plot} we plot the values of Table \ref{tab:rn_meas}.
The red points represent our measurement obtained with the SRT observations.
We fit the spectrum of the northern relic with a continuous injection model (blue line). 
This model takes into account the presence of relativistic particles injected at early stages of the shock passage that have lost their energy resulting in a break in the spectrum.
From the fit we obtain a value of the injection spectral index $\alpha_{\rm inj}$.
This value is related to the spectral index $\alpha_{\rm h}$ as already shown in Eq. \ref{eq:alpha}.\\
From the fit we know that the spectrum breaks at $\nu_{\rm b}=1.8_{-1.5}^{+0.6}$\,GHz and that $\alpha_{\rm inj}=(0.7\pm0.1)$, in agreement with \citet{van10}.
In the high-frequency region of the spectrum (${\rm \nu >> \nu{b}}$), where radiative losses are balanced by the injection of new particles, the resulting spectral index is ${\rm \alpha_h=(1.2\pm 0.1)}$.
By applying the simple DSA model formula:
\begin{equation}
 M = \sqrt{\frac{\alpha_h+1}{\alpha_h-1}}
 \label{eq:dsa}
\end{equation}
we inferred a Mach number ${\rm M=(3.3\pm0.9)}$ in agreement within the errors of the value  ${\rm M=2.7^{+0.7}_{-0.4}}$ presented in \citet{aka2015}.
However, we noticed that there is not yet a clearly detected jump in the X-ray surface brightness as expected at these Mach numbers \citep{donnert17} 
(see discussion in Sect.\ref{sect:ciza}). \\

In the frequency range probed by our observations no significant evidence of steepening in the radio spectrum is found beyond that predicted by the CI model.
From our analysis based on a wide frequency range (from 150 MHz to 8.35\,GHz) we can exclude the steepening found by \cite{stroe2016} beyond 2.5\,GHz. 
Moreover, we find that the continuous injection model is consistent with the data.
Thus, the physics of the northern relic of CJ2242 does not seem 
to require models beyond the standard DSA mechanism, although only new accurate measurements
at frequencies higher than 10\,GHz can definitively exclude alternative scenarios.\\

It should be noted that \cite{stroe2016} presented interferometric measurements at 16\,GHz and 30\,GHz
taken with the Arcminute Microkelvin Imager (AMI) and with the Combined Array for Research in Millimeter-wave Astronomy (CARMA).
These measurements are represented as open dots in Figure \ref{fig:plot}.
We do not include them in our fit procedure 
since these measurements are made with interferometers that could have lost a significant fraction of the flux density from the extended structure.
Nonetheless, we compare them with the best fit of the CI model taking into account the SZ decrement.
Extrapolating the CI model at frequencies higher than 10\,GHz we show as a dot and dashed lines in Figure \ref{fig:plot} the SZ decrement 
as expected by \citet{basu} for this relic assuming a Mach number of M=2.5 and 3.5 respectively. 

As we can see, the decrement is negligible, or at least within the flux calibration uncertainties until $\sim$16\,GHz.
Above this frequency the SZ decrement is increasingly significant, but not enough to explain the measurements at 16 and 30\,GHz.\\
The observed gap could be due to a missing flux problem, as previously mentioned, or it could be real. 
In the latter case a modification from the basic CI model is required.
\citet{donnert} showed that a non uniform magnetic field in the region of the relic could explain the flux densities measured at 16 and 30\,GHz.
Future single-dish observations at frequencies higher than 10\,GHz, which could be obtained with the SRT 7-feed K-band receiver, 
could help to shed light on the claimed high frequency steepening.\\

It should be noted that the break frequency of the CI model corresponds to the spectral break of the oldest electron population injected only if the particles are confined
within the volume of the radio source, and the radio spectrum is extracted from a region that encloses the entire emitting region. 
In this case, we can think at the CI model as the sum of all electron populations from the youngest to the oldest ones.
Below the break frequency (where radiative losses are negligible) particles accumulate and the source luminosity grows linearly with time. 
Above the break frequency a steady state is reached and the high-frequency spectrum stays unchanged since radiative losses are compensated by the freshly injected particles \citep{karda,ci}.
If the magnetic field has been constant in time and uniform in space inside the radio source we can use Eq. \ref{eq:ts} to estimate the time since the start of the injection,
provided that we know the magnetic field strength. 
This assumption was used to estimate the ages of the discrete sources embedded in the diffuse emission in Section \ref{sec:cmeas}.
However, this basic scenario needs to be modified if the confinement time ${\rm \tau_c}$ of the particles inside the source is finite which is likely the case for relics.
We can make the oversimplified assumption that the shocked region consists of a slab of enhanced magnetic field strength of width ${\rm l_{relic}}$.
Particles are accelerated in the outer rim of the slab at the edge of the upstream region by the shock wave and then they flow and age backwards in the downstream region. 
As the particles exit the slab their radio emission rapidly disappears even at low-frequencies due to the drop in the magnetic field strength.
Indeed, the confinement time is related to the relic width by ${\rm \tau_c=l_{relic}/v_d}$, where ${\rm v_d}$ is the downstream velocity. 
Note that a similar argument has been advocated in \citet{carilli} to explain the spectra of the hot spots in Cygnus-A, where particles are injected (or re-accelerated) at the 
termination shock inside the hot spots and then they back-flow into the radio lobes.
By definition, the age of the oldest electrons still present in the relic region and that produce the spectral break seen in the CI spectrum should be exactly $\tau_c$.
The magnetic field strength inside the relic is not known, however, 
from Eq. \ref{eq:ts} it can be shown that the maximum age allowed for the electrons is ${\rm t_{max}=62\,Myr}$, which is obtained for ${\rm B=B_{IC}/\sqrt3\simeq 2.7\,\mu G}$.
We highlight that this value is in agreement with the minimum energy field strength of ${\rm B_{min}=2.4\,\mu G}$ estimated by \citet{maja}.
Therefore, by assuming a  the relic width ${\rm l_{relic} \simeq 200\,kpc}$ \citep{stroe2016}, we can derive a lower limit for the downstream velocity of ${\rm v_d > 3145\,km/s}$ or ${\rm M>3}$.
We note that this lower limit is high but still compatible, within the measurement errors, with the Mach number derived both from the X-rays and from the DSA model (see Eq. \ref{eq:dsa}).
It is reasonable to assume (see Sect. \ref{sect:rm}) that the ambient magnetic field in the proximity of the relic is of the order of 1\,${\rm \mu}$G.
Indeed, the average magnetic field strength inside the relic region
would be amplified by a factor ${\rm \gtrsim 2}$.
This represents the average magnetic field strength in the compression region, while outside we suppose that the magnetic field drops to the background level. 
The particles that flow in the de-compressed region disappear rapidly, even at low frequencies, because of the drop in the spectrum normalization due to the weaker magnetic field.
We stress that, according to this simple scenario, the break frequency of the CI model is not related to the time necessary to the shock to propagate from the cluster centre to the observed position
($\sim$0.5\,Gyr), but rather it refers to the much shorter confinement time of the electrons inside the shocked region.
Therefore, we cannot infer the total age of the relic (that however, we can assume to be larger than ${\rm \tau_c}$)
since the very first injected particles are not contributing anymore to the current observable radio spectrum.
We underline that in this calculation we assumed that the magnetic field strength profile is described by a simple step function.
A more detailed treatment of this issue is beyond the scope of this work; 
we refer to \citet{donnert} for a more sophisticated modelling.
These authors consider much smoother profiles for the magnetic fields after the shock (resulting from their assumption of a small-scale dynamo in the downstream region), 
which results in a curvature for the high-frequency spectrum more pronounced with respect to that of the CI model.\\
In any case, it is important to point out that a steep-spectrum power-law behaviour is expected only if the confinement time is so long that the break frequency of the oldest population present in the relic region 
shifts below the lower end of the observed window.
Attempts to fit the spectrum with a single power-law have been presented in \cite{maja}. 
They found for CJ2242 ${\rm\alpha=0.90\pm0.04}$ which is intermediate between ${\rm\alpha_{inj}}$ and ${\rm\alpha_h}$ in our fit.
However, this value of spectral index leads to problematic results since the Mach number expected by the DSA model diverges.
We therefore opt for the CI fit, under the assumption that the spectral curvature is real, since the missing flux problem is negligible below 1\,GHz.
This is confirmed by the finding that the flux density of the relic before and after the single-dish-interferometric combination is roughly the same at L-band.\\

Concerning the southern relic: we fitted our estimates of the flux density at 1.4\,GHz and 6.6\,GHz with a simple power-law as shown in Figure \ref{fig:sr}.
Because only two experimental points were available the fit of the CI model is not applicable.
The southern relic has a spectral index $\alpha\simeq1.9$ corresponding to a Mach number of $M\simeq 1.8$.
Even in this case the value is in agreement with that found by \citet{aka2015} ($M=1.7^{+0.4}_{-0.3}$).\\

After the submission of our paper we became aware of the work of \citet{lofar} where these authors present Low-Frequency Array (LOFAR) observations between 115.5 and 179\,MHz 
and a study of the diffuse sources 
of CJ2242. Assuming the DSA model and by evaluating the injection spectral index in the upstream region,
they find Mach numbers for the southern and the northern relics in agreement with the ones obtained from X-rays by \citet{aka2015} and consistent 
with our results.

\begin{figure*}\centering
 \includegraphics[scale=0.9]{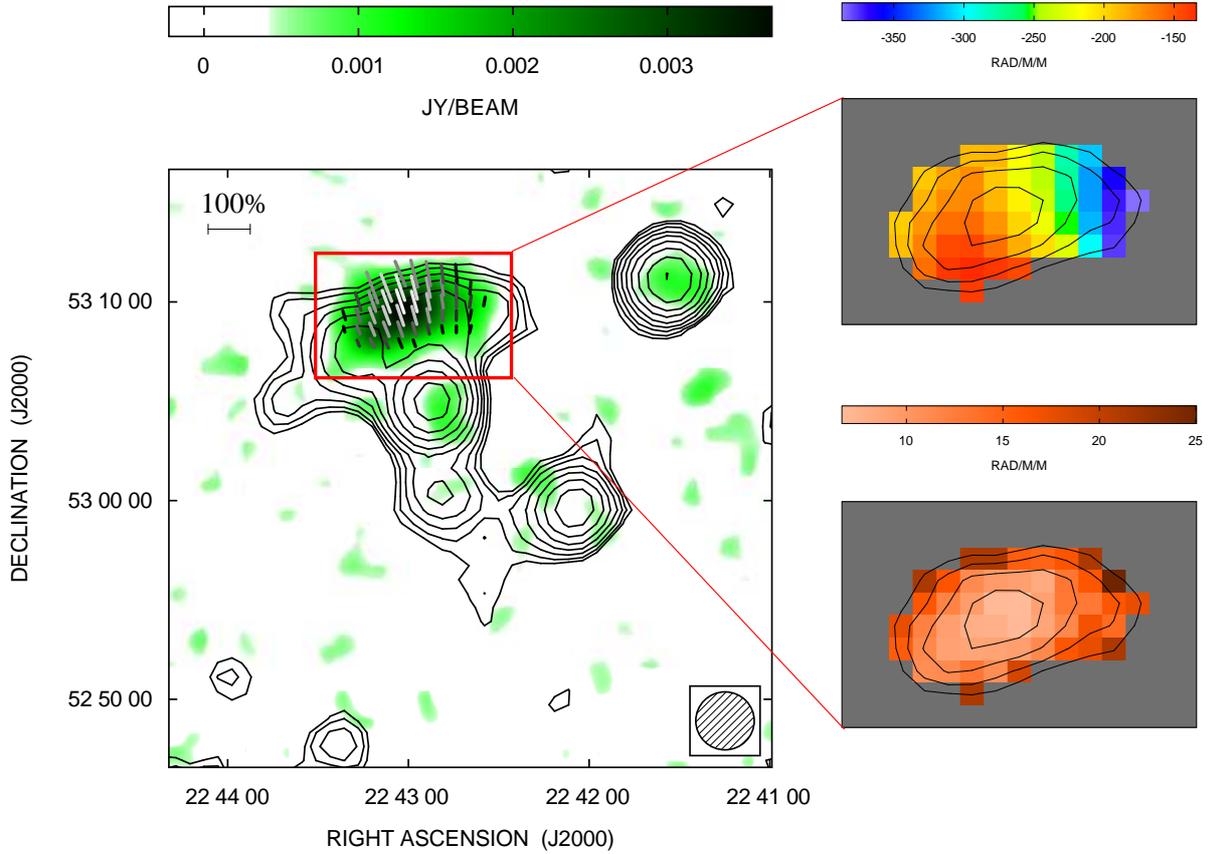}
 \caption{{\it Left:} SRT linearly-polarized intensity image of the galaxy cluster CIZA J2242.8+5301 obtained from the spectral average in the frequency range 6-7.2\,GHz. 
 The field of view of the image is 30$^{\prime}\times30^{\prime}$. 
 The FWHM beam is 2.9$^{\prime}$ and is shown in bottom-right corner.
 The noise level is 0.35 mJy/beam.
 Contours refer to the total intensity image starting at the 3${\rm\sigma}$-level and increasing by factors of ${\rm\sqrt{2}}$.
 The vectors indicate the electric field polarization with the orientation corresponding to the polarization angle and the length proportional to the 
 polarization percentage where 100$\%$ is shown in the top-left corner. 
 The vectors are traced only for pixels with a total and linearly-polarized intensity signal higher than 3${\rm\sigma}$ and an error of the polarization angle below 10$^{\circ}$. 
 {\it Right:} Result of the application of the RM Synthesis technique. The top panel on the right represents the Faraday depth we obtained for the maximum polarized signal.
 In the bottom panel we report the associated uncertainties evaluated from simulations.}
 \label{fig:pol_deep}
\end{figure*}
\section{Polarized intensity results}
\subsection{C-band}
Figure \ref{fig:pol_deep} shows, on the left,
the SRT 6.6\,GHz linearly-polarized intensity image of CJ2242 in colours as detected from the deep observations (FoV=$30^{\prime}\times30^{\prime}$).
Contours refer to the total intensity image (Figure \ref{fig:i_deep}) and vectors indicate the electric-field polarization.
In this image the noise-level is ${\rm\sigma}$=0.35 mJy/beam.\\
We detect the polarized emission of the northern relic: its average fractional polarization is of $\sim$40$\%$, reaching levels as high as $\sim$70$\%$.
This could be a consequence of the shock passage - the turbulent magnetic field is compressed and its component in the direction of the shock is suppressed.

\subsection{Rotation Measure Synthesis}
\label{sect:rm}
We applied the Rotation Measure (RM) Synthesis technique \citep{burn,brent} on the linearly-polarized data presented in Figure \ref{fig:pol_deep} in order to recover the Faraday depth of the relic.\\
In addition to the SRT data, for which the frequency band goes from 6.0\,GHz up to 7.2\,GHz with a channel width $\sim$1.46 MHz, 
we also included two more channels taken from the archival VLA data at 4.835 and 4.885\,GHz presented in Sect. \ref{sec:cmeas}.
This helped us to improve the reliability of the RM Synthesis technique by extending the ${\rm \lambda^2}$ coverage.
We smoothed the VLA U and Q images to the SRT resolution.
However, the resulting RM transfer function has still a quite large FWHM ($\sim$4000\,rad/m$^2$) since we are observing at relatively high frequencies.
We should also consider that the angular resolution of our polarized images corresponds to a linear scale of $\sim$550 kpc, which could be larger than the autocorrelation length of the RM
fluctuations.
Indeed, it is hard to distinguish multiple components in the Faraday depth space and as a consequence we cannot verify if the RM originates internally to the radio relic or in an external
Faraday screen.
For these reasons, in what follows we assumed that the RM originates entirely in the ICM between us and the relic and that the internal Faraday rotation is negligible.
In practice we considered just one polarized component in Faraday space and we computed its RM from the RM Synthesis by measuring the Faraday depth of the polarization peak.

Figure \ref{fig:pol_deep} shows, on the right, the results of the RM Synthesis applied on pixels with a polarized S/N ratio larger than 3.
The top-right panel shows the resulting Faraday depth while the bottom-right panel shows the associated uncertainty.\\
These errors have been evaluated with a Monte Carlo simulation.
We injected 30 different components in Faraday depth, randomly distributed between $-$10000 and 10000\,rad/m$^2$, 
with intensities ranging from 1\,mJy/beam up to 4\,mJy/beam to reproduce the S/N range of the observed polarization intensity.  
From these values we produced simulated U and Q data in the selected frequency band (SRT C-band plus the two VLA frequencies). 
We fixed the relic spectral index to what found in Sect. \ref{sect:nr_spix} and we assumed a weight $w_{\rm ch}$ for each frequency channel (defined as $w_{\rm ch}=1/(\sigma_{\rm ch}^2)$
where $\sigma_{\rm ch}$ is the rms of the single channel), in order to accurately reproduce the effect of the noise on data. 
We used $w_{\rm ch}=0$ for those channels that have been blanked due to RFI.  
We applied the RM Synthesis to the simulated U and Q data and we compared the injected values with respect to the polarized signal and the Faraday depth inferred from the RM Synthesis.
We assumed for the RM uncertainty the rms of the distribution of the difference between input and ``measured'' Faraday depth.
As expected, the higher the S/N ratio the lower the rms.
For a S/N=3 the RM uncertainty is about 25 rad/m$^2$, but it decreases down to 7 rad/m$^2$ at the position with the brightest polarized emission (S/N=10).

The Faraday depth image shows negative values with a gradient along the relic length from about $-$150 at the east end to about $-$400 rad/m$^2$ at the west end.
At the polarization peak we found ${\rm RM=(-176\pm8)\,rad/m^2}$.
Our Faraday depth image is in good agreement with the values obtained by \citet{maja},
but we observe a RM gradient of $\sim$250\,rad/m$^2$ over an angular scale of 10$^{\prime}$, 
a factor of 2 larger than the gradient inferred from that work. \\

We estimated the Galactic RM contribution in the position of CJ2242 using the reconstruction of \citet{op} that provides a ${\rm RM_{gal}=(-73\pm70)\,rad/m^2}$.
Therefore, despite the large uncertainties the Faraday depth shown in Figure \ref{fig:pol_deep} seems to exceed the Galactic contribution by at least a factor of two.
Furthermore one could ask if 
the observed RM gradient is due to a gradient in the foreground Galactic RM since CJ2242 is very close to the Galactic plane (latitude b=$-$5.11 deg).
Using the map of \citet{op} we deduced that in general RM gradients as large as 1000\,rad/m$^2$/deg are possible along the Galactic plane.
In order to explain the RM gradient observed along the northern relic in CJ2242 we would need a very high gradient of about 1500\,rad/m$^2$/deg.

We therefore deduce that the RM gradient is not due to the Galactic foreground.
Thus, we can subtract the constant value of ${\rm RM_{gal}=-73\,rad/m^2}$, and evaluate if the residual RM could be caused by the magneto-ionic medium in the galaxy cluster itself.
Following \citet{fede10}, and
assuming a cluster magnetic field with an autocorrelation length ${\rm \Lambda_B<<L}$, where L is the integration path along the gas density distribution, 
the observed RM along a line of sight is a random walk process that involves a large number of cells with size ${\rm \Lambda_B}$.
The distribution of the RM will be a Gaussian with zero mean and a variance:
\begin{equation}
 \sigma_{RM}^2=<RM^2>=(812)^2 \Lambda_B \int_0^L (n_e B_{\parallel})^2 dl
\end{equation}
Assuming that the cluster magnetic field follows a $\eta$-profile with respect to the density distribution $B(r)=B_0(n_e(r)/n_0)^{\eta}$ 
and that the density distribution follows a $\beta$-profile $n(r)=n_0(1+r^2/r_{\perp}^2)^{-3\beta/2}$, we obtain that the RM dispersion at a given projected distance $r_{\perp}$ is:
\begin{equation}
 \sigma_{RM}(r_{\perp},L)= \frac{K(L) B_0 \Lambda_B^{1/2} n_0 r_c^{1/2}}{\large(1+\frac{r_{\perp}^2}{r_c^2}\large)^{\frac{6\beta(1+\eta)-1}{4}}} \sqrt{\frac{\Gamma[3\beta(1+\eta)-\frac{1}{2}]}{\Gamma[3\beta(1+\eta)]}}
\end{equation}
where K(L) depends on the integration path along the gas density distribution. We assume that the northern relic is located halfway the cluster so that K=441. 
${\rm B_0}$ and ${\rm n_0}$ are respectively the central values of the magnetic field and thermal plasma density profiles,
${\rm r_c}$ is the core radius of the cluster and ${\rm \Gamma}$ is the gamma function.
The observed RM image is compatible with a magnetic field tangled on a scale equal or larger than the RM structure of Figure \ref{fig:pol_deep}, so that we can assume 
${\rm \Lambda_B\sim1300\,}$kpc. 
If we subtract from the RM of the peak the Galactic contribution the residual RM due to the ICM is ${\rm RM=-103\,rad/m^2}$.
Assuming reasonable values for the parameters of the hot ICM (${\rm n_0=10^{-3}\,{cm}^{-3}}$, ${\rm\beta=0.6}$, and ${\rm r_c=500\,}$kpc),
we would need a magnetic field of ${\rm B_0=4.5\,\mu G}$ at the cluster centre with ${\rm \eta=0.5}$ to generate
an RM distribution with a ${\rm \sigma_{RM}\sim107\,rad/m^2}$ at the relic, where we expect ${\rm B=1.6\,\mu}$G.
We note however that even if the central magnetic field strength is in line with what typically found at the centre of galaxy clusters \citep{feretti12},
the magnetic field autocorrelation length is much larger than what expected for the turbulent ICM.
We also note that the condition ${\rm \Lambda_B<<L}$ could not be verified in our case, since the scale of the magnetic field would be comparable with the size of the cluster.\\

Another possibility is to consider that the Faraday rotation is occurring in a cosmic web filament which includes CJ2242.
As suggested by \citet{planck} the
primordial magnetic fields have a strength not larger than a few nG.
Even assuming a thermal density of ${\rm n_{e}=10^{-4}\,cm^{-3}}$ and a magnetic field tangled on a scale of 1300 kpc, the resulting RM would be only $\sim$0.1\,rad/m$^2$,
making the filament contribution irrelevant.
Moreover we notice that magnetic fields of the order of ${\rm\sim0.1\,\mu}$G in filaments should be reached only in the presence of small-scale dynamo amplification in excess of what can be presently resolved
by simulation \citep{franco}.
This may be verified in the presence of small-scale vorticity and/or plasma instabilities of various kinds \citep[e.g.][]{mogavero}, which would however make the 
magnetic field tangled on much smaller scales than what is inferred by our observations.\\

For this analysis it is clear that high resolution polarized observations are needed to clarify the nature of the observed Faraday rotation in CJ2242.
For instance the forthcoming Westerbork Observations of the Deep APERTIF Northern-Sky \citep[WODAN,][]{wodan} project will give us a polarized image of the northern sky.
The high resolution and sensitivity of this survey will make it possible to better investigate the RM observed in the region covered by the galaxy cluster CJ2242
and will help to constrain the properties of the intracluster magnetic field. 

\section{Conclusions}
   We observed the galaxy cluster CIZA J2242.8+5301 with the Sardinia Radio Telescope to further study its diffuse radio emission.
   We conducted observations in three frequency bands centred at 1.4\,GHz, 6.6\,GHz and 19\,GHz.
   These single-dish data were also combined with archival interferometric observations at 1.4 and 1.7\,GHz taken with WSRT. \\

   From the single-dish-interferometer combined images we measured a flux density of ${\rm S_{1.4GHz}=(158.3\pm9.6)\,mJy}$ for the central radio halo and
   ${\rm S_{1.4GHz}=(131\pm8)\,mJy}$ and ${\rm S_{1.4GHz}=(11.7\pm0.7)\,mJy}$ for the northern and the southern relics respectively.
   At 6.6\,GHz we measured ${\rm S_{6.6GHz}=(19.3\pm1.1)\,mJy}$ for the northern relic and ${\rm S_{6.6GHz}=(0.9\pm0.5)\,mJy}$ for the southern relic. 
   Assuming simple diffusive shock acceleration we interpret the measurements of the northern relic with a continuous injection model, 
   represented by a broken power-law. 
   This gave us an injection spectral index ${\rm\alpha_{inj}=0.7\pm0.1}$ and a spectral index ${\rm\alpha=1.2\pm0.1}$, resulting in a Mach number ${\rm M=3.3\pm0.9}$, 
   consistent with the recent X-ray estimates.
   No significant steepening of the relic radio emission beyond 2.5\,GHz is seen in the data up to 8.35\,GHz.
   By fitting with a simple power-law spectrum (${\rm S_{\nu}\propto \nu^{-\alpha}}$) the measurements of the southern relic, we obtained a spectral index ${\rm \alpha=1.7\pm 0.7}$,
   corresponding to a Mach number ${\rm M=2.0\pm0.7}$, in agreement within the errors with the X-ray estimates.
   The properties of the radio halo, namely its largest linear size and radio power, together with the cluster X-ray luminosity, 
   have been compared to those of the other radio haloes known in literature. 
   We found that the radio halo in CJ2242 is the most largest and among the most luminous haloes ever studied.\\
   
   In the SRT image at 1.4\,GHz we noticed an extended ``L-shaped'' emission located about 10\,Mpc North-East of CJ2242. 
   The radio emission seems to connect a few spots of X-ray emission.
   In particular the X-ray source closest to CJ2242 is associated with the RASS Faint Source Catalogue source 1RXS J224504.3+532800, 
   which has a hardness ratio of 0.93$\pm$0.09, and with the second RASS Source Catalogue source 2RXS J224454.9+532719 with a hardness ratio of 1.0$\pm$0.1.
   Because of its high hardness ratio, 
   we classify this source as a galaxy cluster candidate.
   Moreover at the centre of the X-ray source we found a head-tail radio galaxy that further supports the presence of a dense medium, typical of a galaxy cluster.
   The head-tail is associated with the galaxy 2MASX J22445464+5328318 and at less than 0.7$^{\prime}$ we also found the galaxy 2MASX J22445565+5327578. 
   No spectroscopic redshifts are available for these galaxies, however,
   by using the K-band magnitude-redshift relations of \citet{inskip} we estimated a redshift z$\sim$0.1$-$0.2 close to that of CJ2242.
   We obtained a similar result (z$\sim$0.15$-$0.2) when considering the K-band magnitude-redshift relation for the brightest cluster galaxies \citep{stott}.\\
   
   With the help of the new SRT measurements at 19\,GHz
   we studied the spectra of the radio galaxies in CJ2242 with the aim to subtract their contribution to the extended emission of the 6.6\,GHz SRT image.
   In addition, we estimated the synchrotron age of these radio galaxies. We found that most of them are active sources whose radio spectrum is very well
   reproduced by a continuous injection model. 
   However, one of them we classify as a dying radio galaxy on the basis of the exponential cut-off in the integrated spectrum and 
   on the relaxed morphology.
   This finding confirms the tendency for these rare radio sources to be preferentially found in the dense environment of galaxy clusters \citep{murgia11}.
   We notice that the dying source is close to an arc-like feature
   which appears to be a secondary shock or a possible extension of the northern relic, although a discontinuity between these two structures is present.
   The remnant lobes could be a source of seed electrons for the relic shock-wave, a possibility that has already been suggested in the case of other galaxy clusters 
   \citep{bonafede14,van17}.\\

   Finally, we evaluated the rotation measure of the northern relic by applying the RM Synthesis technique at the 6.6\,GHz data.
   The Faraday depth image shows negative values with a gradient along the relic length from about $-$150 at the east end to about ${\rm-400\,rad/m^2}$ at the west end.
   At the polarization peak we found RM=${\rm(-176\pm8)\,rad/m^2}$.
   Our Faraday depth image is in good agreement with that obtained by \citet{maja}.
   We derive the presence of a RM gradient of ${\rm\sim250\,rad/m^2}$ over an angular scale of 10$^{\prime}$, 
   a factor of 2 larger than the gradient inferred in that work.
   These results provide insights on the magnetic field structure of the ICM surrounding the relic,
   but further observations are needed to clarify the nature of the observed Faraday rotation.\\

   In conclusion, this study demonstrates that single dish observations can be helpful to properly study the diffuse emission in galaxy clusters, 
   especially when used in combination with interferometric observation at higher resolution. 
   Future observations at frequencies higher than 10\,GHz could be obtained with the new SRT 7-feed K-band receiver, 
   and could play a decisive role to constraining the physics of relic sources, and helping to distinguish between the different acceleration models proposed 
   to explain the origin of the relativistic electrons. 

\section*{Acknowledgements}
We thank an anonymous referee for her/his comments and suggestions which helped us to improve this paper.
The Sardinia Radio Telescope \citep{bolli,scicom} is funded by the Ministry of Education, University and Research (MIUR), Italian Space Agency (ASI), the Autonomous Region of Sardinia (RAS) and INAF itself 
and is operated as National Facility by the National Institute for Astrophysics (INAF).
The development of the SARDARA back-end has been funded by the Autonomous Region of Sardinia (RAS) 
using resources from the Regional Law 7/2007 \textquotedblleft Promotion of the scientific research and technological innovation in Sardinia\textquotedblright \, 
in the context of the research project
CRP-49231 (year 2011, PI Possenti): \textquotedblleft High resolution sampling of
the Universe in the radio band: an unprecedented instrument to understand the fundamental laws of the nature\textquotedblright.
F. Loi gratefully acknowledges Sardinia Regional Government for the financial support of her PhD scholarship 
(P.O.R. Sardegna F.S.E. Operational Programme of the Autonomous Region of Sardinia, European Social Fund 2007-2013 - Axis IV Human Resources, Objective l.3, Line of Activity l.3.1.).
This research was partially supported by PRIN-INAF 2014.
The National Radio Astronomy Observatory (NRAO) is a facility of the National Science Foundation, operated under cooperative agreement by Associated Universities, Inc. This research made use of the
NASA/IPAC Extragalactic Database (NED) which is operated by the Jet Propulsion Laboratory, California Institute of Technology, under contract with the National Aeronautics and Space Administration.
Basic research in radio astronomy at the Naval Research Laboratory is funded by 6.1 Base funding.
This research was supported by the DFG Forschengruppe 1254 Magnetisation of Interstellar and Intergalactic Media: The Prospects of Low-Frequency Radio Observations.
F. Vazza acknowledges funding from the European Union's Horizon 2020 research and innovation programme under the Marie-Sklodowska-Curie grant agreement No 664931.
This publication makes use of data products from the Two Micron All Sky Survey, which is a joint project of the University of Massachusetts and the Infrared Processing and Analysis Center/California Institute of Technology, 
funded by the National Aeronautics and Space Administration and the National Science Foundation.

\bsp	
\label{lastpage}
\end{document}